\documentclass[twocolumn,aps,prd,longbibliography, replace]{revtex4-1}

\usepackage{graphicx,bm,xcolor, bbold}
\usepackage[normalem]{ulem}
\usepackage{hyperref}
\hypersetup{colorlinks, linkcolor={blue},citecolor={blue},urlcolor={blue}}
\usepackage{physics}
\usepackage[utf8]{inputenc}
\usepackage[english]{babel}
\usepackage{tikz}

\makeatletter
\def\graphicscale{\twocolumn@sw{0.3}{0.4}}
\def\graphicthreescale{\twocolumn@sw{0.3}{0.4}}

\begin{document}
\title{Liouvillian gap and out-of-equilibrium dynamics of a sunburst Kitaev ring:\\ from local to uniform dissipation}

\author{Alessio Franchi}
\affiliation{Dipartimento di Fisica dell'Universit\`a di Pisa
        and INFN, Largo Pontecorvo 3, I-56127 Pisa, Italy}

\author{Francesco Tarantelli}
\affiliation{Dipartimento di Fisica dell'Universit\`a di Pisa
        and INFN, Largo Pontecorvo 3, I-56127 Pisa, Italy}

\date{May 2, 2023}

\begin{abstract}
We consider an open quantum system composed of a $(1+1)$-dimensional Kitaev ring coupled with the environment via $n$ particle-loss dissipators in a \textit{sunburst} geometry. We describe the out-of-equilibrium dynamics of the whole apparatus in terms of Lindblad master equations and focus on the scaling behavior of the Liouvillian gap $\Delta_\lambda$ with the system size $L$. We unveil different regimes, which depend on the number of dissipation sources considered in the large-size limit and the dissipation strength $w$, which can be either fixed or attenuated to zero as $w\sim1/L$. In the second part, we develop a dynamic Finite-Size Scaling framework close to Continuous Quantum Transitions to monitor the time evolution of the critical correlations and the entanglement entropy, emphasizing the role of $\Delta_\lambda$ in this regime.   
\end{abstract}
\maketitle
\section{Introduction}

The progress achieved in the control of nano-scales many-body systems has recently renewed the interest in understanding the out-of-equilibrium dynamic in quantum spin models~\cite{PSSV-2011-noneqcoll, GAN-2014-quantumsimulation}. A deeper comprehension of the time evolution of the critical correlations and entanglement spreading is indeed sought by both the theoretical and experimental communities~\cite{ADM-2015-EntanglementReview}. Since any experimental device is unintentionally coupled to the environment, a particular emphasis is put on the dynamics of \textit{open quantum systems}~\cite{BP-openquantumsystembook}.

When the interactions of a quantum system with its surroundings are sufficiently weak, the real-time evolution of such apparatuses emerges from the interplay between the unitary and dissipative dynamics of the whole setup~\cite{RV-2021-coherentanddissipativedynamicsreview}. These hypotheses are usually satisfied within \textit{Lindblad} frameworks, which underpin the modelization of most atomic, molecular, and optical devices (AMO)~\cite{BDS-2015-KeldyshOptical}. In such cases, the system is described in terms of a density matrix $\rho$, and the time evolution is controlled by \textit{Linblad Master equations}
\begin{equation}
    \frac{d\rho}{dt} = \mathcal{L}[\rho]\,.
    \label{eq_def_intro_lindblad}
\end{equation}
The system generally thermalizes to a Non-Equilibrium Steady-State (NESS) solution after a transitory time frame. However, determining whether the NESS is unique is a more subtle issue~\cite{N-2019-uniquenesslindblad, SW-2010-openuniquesolution}. A quantity of particular interest is the \textit{Liouvillian gap}, hereafter denoted as $\Delta_\lambda$. This energy scale sets the typical relaxation time required to make the NESS stand out, entailing a complete loss of information on the initial quantum state. Quantum memory devices, for example, would benefit from long relaxation times, therefore small $\Delta_\lambda$~\cite{CCP-2011-quantummemories}. 

Several works have addressed the nature of the Liouvillian gap in one-dimensional open quantum systems, considering different lattice geometries and dissipation sources also in integrable models~\cite{Z-2015-relaxtimes}. Distinguished behaviors emerge when the dissipators are either isolated or in a relatively large number compared to the system size $L$.
On the one hand, with bulk dissipation acting on the whole network, the system is gapped in several paradigmatic spin chains, such as XX, XXZ, and Ising models~\cite{YWHWD-2021-artificialnetweork, Z-2015-relaxtimes, KS-2019-nonhermitiankitaevladder}. On the other hand, when the number of dissipative sources is constant, the Liouvillian gap generally vanishes with a distinctive power-law behavior in the thermodynamic limit, typically as $\sim L^{-3}$~\cite{KS-2020-boundarydephasing, TV-2021-dissipativeboundaries, Z-2011-XXXchaingap}.
The physical mechanisms tying together these two regimes are still unclear and are the main focus of this work.

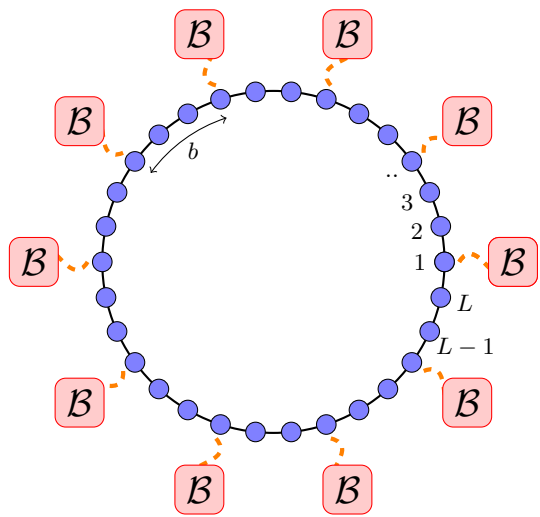
\begin{figure}
    \centering
\begin{tikzpicture}[scale=0.65]
    \draw[thick] (0,0) circle (3.5cm);

    \foreach \x in {0,12,...,360}
    \filldraw [fill=blue!50] (\x:3.5cm) circle (0.2);

    \foreach \x in {0,36,...,360}
    \draw[orange, dashed, ultra thick] (\x:3.8cm) .. controls (\x+8:4.2cm) and (\x-8:4.6cm) .. (\x:4.6cm);

    \foreach \x in {0,36,...,360}
    \node[rectangle,
    draw = red,
    text = black,
    fill = red!20!white,
    rounded corners,
    minimum size=0.65cm] (r) at (\x:4.9cm) {\Large $\mathcal{B}$};

    \draw[<->] (108:3.1cm) arc
    [start angle=108,
        end angle=144,
        radius=3.1cm,
    ] ;

    \node[] (s1) at (126:2.8cm) {$b$};

    \node[] (s1) at (0:3cm) {$1$};
    \node[] (s2) at (12:3cm) {$2$};
    \node[] (s3) at (24:3cm) {$3$};
    \node[] (s4) at (36:3cm) {$..$};
    \node[] (s5) at (-12:4cm) {$L$};
    \node[] (s6) at (-24:4.3cm) {$L-1$};
\end{tikzpicture}
    \caption{Sketch of a Kitaev ring with $L=30$ qubits coupled with $n=10$ dissipators in a \textit{sunburst} geometry ($b=3$ in the figure).}
    \label{fig_sketch_sunburst_dissipation}
\end{figure}

We then consider a lattice model tailored to unveil the crossover regime between the  dissipation schemes presented. We investigate a $(1+1)$-dimensional Kitaev ring with local particle-decay dissipators arranged in a \textit{sunburst} geometry~\cite{FRV-staticsunburst, FRV-timesunburst, MS-2022-sunburstquench}. The whole apparatus is sketched in Fig.~\ref{fig_sketch_sunburst_dissipation}. The open quantum system is coupled with the environment by means of $n\equiv L/b$ equally-spaced external baths, which reduce to some extent the translation invariance of the starting model. We explore different large-size limits, depending on the number of dissipators taken into account. A thorough study of the Liouvillian gap $\Delta_\lambda$ is the main focus of the first part of this paper. In the second part, we examine the real-time evolution of the system, triggered by a \textit{soft quench} of a coupling constant appearing in the defining hamiltonian~\footnote{In a \textit{soft quench} the variation of the quenched parameter is attenuated down to $0$ with increasing the lattice size $L$.}. Starting the protocol in the proximity of a Continuous Quantum Transition (CQT), we study the out-of-equilibrium dynamic using Renormalization Group (RG) arguments and Finite-Size Scaling (FSS) frameworks~\cite{C-1996-ScalingandRG, RV-2021-coherentanddissipativedynamicsreview}. We emphasize the interplay between the unitary and dissipative dynamics and the role played by the gap $\Delta_\lambda$, extending some of the results already presented in Ref.~\cite{NRV-2019-competingdissipativeandcoherent} to our model. To outline our FSS theory, we mainly focus on the scaling properties of the critical correlations and one of the most common entanglement quantifiers, i.e., the entanglement entropy~\cite{ZMZ-2021-Renyientropiesopen}.

The paper is organized as follows. In Sec.~\ref{sec_introduce_the_model}, we present the model in detail. In Sec.~\ref{sec_liouville gap}, we address the scaling properties of the Liouvillian gap $\Delta_\lambda$, considering either the cases with a fixed or an increasing number of dissipation sources in the large-size limit. Sec.~\ref{sec_fss} is devoted to the development of the out-of-equilibrium FSS framework in the presence of dissipation mechanisms at CQTs. We finally draw our conclusion and present future outlooks in Sec.~\ref{sec_conclusions}.

\section{The lattice model}
\label{sec_introduce_the_model}
We study a fermionic Kitaev ring in one spatial dimension with antiperiodic boundary conditions (APBC), therefore considering $\hat{c}_{L+1}=-\hat{c}_1$. The hamiltonian follows
\begin{equation}
\begin{aligned}
    \hat{H}&=-\sum_{x=1}^{L} (\hat{c}^{\dagger}_{x}\hat{c}_{x+1} + \hat{c}^\dagger_{x}\hat{c}^\dagger_{x+1}+h.c.) - \mu\sum_{x=1}^{L} \hat{n}_x\,,
    \label{eq_kitaev}
\end{aligned}
\end{equation}
where $\hat{n}_x\equiv\hat{c}^\dagger_x\hat{c}_x$ is the number operator on the site $x$, and the operators $\hat{c}_x, \hat{c}^\dagger_x$ satisfy the canonical anticommutation relations, thus $\{\hat{c}_x, \hat{c}_y\}=\{\hat{c}^\dagger_x, \hat{c}^\dagger_y\}=0$ and $\{\hat{c}_x, \hat{c}^\dagger_y\}=\delta_{xy}$. Applying the Jordan-Wigner transformation~\cite{S-1999-QuantumBook}, the Kitaev ring can be exactly mapped into a quantum Ising chain with a transverse field~\cite{P-1970-Isingmodel}. We point out that the transformation does not preserve also the same boundary conditions, so attention should be paid when recasting Eq.~\eqref{eq_kitaev} in its bosonic counterpart~\cite{RV-2021-coherentanddissipativedynamicsreview}. Nonetheless, many bulk properties of the Ising model, such as the critical exponents at the Quantum Critical Point (QCP), are preserved by the mapping.

The quantum Ising model with a transverse field is one of the most common theoretical laboratories where fundamental issues on quantum phase transition can be addressed, given our deep knowledge of the FSS properties and quantum correlations~\cite{S-1999-QuantumBook}. The model is characterized by a $\mathbb{Z}_2$ global symmetry under spin reflection along the longitudinal axis. In Eq.~\eqref{eq_kitaev}, this symmetry is implemented by the transformation that maps $\hat{c}^{(\dagger)}_x\to -\hat{c}^{(\dagger)}_x$. At zero temperature, the ground state experiences a CQT in the same universality class as the two-dimensional Ising model (it takes place at $\mu_c=-2$ in our notations), and the $\mathbb{Z}_2$ symmetry is then spontaneously broken. The critical point separates a paramagnetic phase ($\abs{\mu}<\abs{\mu_c}$), where correlation functions are exponentially dumped, from an ordered phase ($\abs{\mu}<\abs{\mu_c}$), where correlation functions are instead long-range ordered. Close to the critical point, the correlation length diverges as $\xi\sim\abs{\mu-\mu_c}^{-\nu}$, where $\nu=1/y_g=1$ for Ising transitions. The gap $\Delta$, which describes the energy difference between the first excited state and the ground state, vanishes instead as $\Delta\sim\xi^{-z}$ with $z=1$. 

To model the weak interaction between the \textit{open} quantum ring and the surrounding environment, we consider $n=L/b$ uniformly spaced local operators that are always commensurate to the number of sites of the chain $L$. The whole setup is then put in a \textit{sunburst} geometry, see also Fig.~\ref{fig_sketch_sunburst_dissipation}, and translation invariance is narrowed down to translations that are multiples of $b$~\cite{FRV-staticsunburst, FRV-timesunburst, MS-2022-sunburstquench}. We work under the Born-Markov and secular approximations, so dissipators can be effectively modeled employing Lindblad \textit{quantum jump} operators $\hat{L}_x$. In this limit, the time evolution of the density matrix can be described by Markovian master equations in the Lindblad form as~\cite{RV-2021-coherentanddissipativedynamicsreview, BP-openquantumsystembook} 
\begin{equation}
    \frac{d\rho}{dt}=\mathcal{L}[\rho]\equiv-i\big[\hat{H},{\rho}\big]+\mathbb{D}[\rho]\,,
    \label{eq_def_liouvillian}
\end{equation}
where $\mathcal{L}$ is the Liouville superoperator, and $\mathbb{D}$ is the corresponding dissipation term, whose strength is regulated by the homogeneous coupling $w$
\begin{equation}
\begin{aligned}
    \mathbb{D}[\rho]&=w\sum_{x=0}^{n-1}\mathbb{D}_{xb+1}[\rho]\,,\\
    \mathbb{D}_{x}[\rho]&=\hat{L}_{x}\rho\hat{L}^\dagger_{x}-\frac{1}{2}\big\{\hat{L}^\dagger_{x}\hat{L}_{x},\rho\big\}\,.
\end{aligned}
\label{eq_def_dissipator}
\end{equation}
In this work, we focus on the case of particle-decay jump operators, i.e., $\hat{L}_x=\hat{c}_x$, where fermionic particles are continuously removed from the site $x$. With this choice, the Liouville operator $\mathcal{L}$ is quadratic in the fermionic variables $\hat{c}_x$ and $\hat{c}^\dagger_x$, and, in this sense, we say that the open ring we study maintains its integrability. Most of the results discussed in this work should preserve their validity also for particle-pumping dissipation ($\hat{L}_{x}=\hat{c}^\dagger_x$), since Eq.~\eqref{eq_def_liouvillian} is still quadratic in the fermionic creation and annihilation operators.

\section{Liouvillian gap}
\label{sec_liouville gap}

 This section is devoted to discussing the different scaling behaviors observed for the Liouvillian gap $\Delta_\lambda$. As mentioned in the introduction, we will consider two different limits, depending on the number of dissipation sources considered with increasing the lattice size. We first review some useful definitions related to the Liouvillian gap after rephrasing Eq.~\eqref{eq_def_liouvillian} into a standard eigenvalue problem. To this purpose, let us consider the following equation
\begin{equation}
    \widetilde{\mathcal{L}}[\widetilde{\rho}_i]=\lambda_i \widetilde{\rho}_i\,, \quad \lambda_i\in\mathbb{C}\,,
    \label{eq_eigenvalue_lindbladian}
\end{equation}
where $\widetilde{\mathcal{L}}$ is the (non-hermitian) Lindblad superoperator derived from Eq.~(\ref{eq_def_liouvillian}) after the Choi-Jamiolkowski isomorphism~\cite{VZ-2004-superoperatorvidal, BP-openquantumsystembook}, and $\widetilde{\rho}_i$ is the density matrix eigenoperator associated with the complex eigenvalue $\lambda_i$. In a few words, the transformation we have mentioned sends the density matrix $\rho$ to $\widetilde{\rho}$ through the mapping $\rho_{ij}\ket{i}\bra{j}\to\widetilde{\rho}_{ij}\ket{i}\ket{j}$. Therefore, the vectorized $\widetilde{\rho}$ lives in a $4^{L}$-dimensional Hilbert space. In this basis, the action of $\widetilde{\mathcal{L}}$ on $\widetilde{\rho}$ can be written as follows
\begin{equation}
\begin{aligned}
\widetilde{\mathcal{L}} =& -i \big(\hat{H} \otimes \hat{\mathbb{1}} - \hat{\mathbb{1}}\otimes \hat{H}^t \big) + w\sum_{x=0}^{n-1}\hat{L}_{bx+1}\otimes \hat{L}^*_{bx+1}\\
&-\frac{w}{2}\sum_{x=0}^{n-1}\big(\hat{L}^{\dagger}_{bx+1}\hat{L}_{bx+1}\otimes\hat{\mathbb{1}}+\hat{\mathbb{1}}\otimes\hat{L}^t_{bx+1}\hat{L}^*_{bx+1}\big)\,.
\end{aligned}
    \label{eq_widetilde_mathcal_L_def}
\end{equation}
It can be shown that all eigenvalues of $\widetilde{\mathcal{L}}$ satisfy $\Re{\lambda_i}\leq0$~\cite{BP-openquantumsystembook}. The zero mode of the above operator represents the steady-state solution, namely, the NESS of the system. Since $\hat{L}_x$ is not hermitian in the case of particle-decay dissipation, the density matrix corresponding to the steady-state solution is not proportional to the identity matrix~\cite{KS-2020-boundarydephasing}. We characterize completely the asymptotic state for the simplest case $b=1$, describing its features and proving its uniqueness in App.~\ref{sec_steadystate}. For $b>1$, our numerical data have not highlighted any relevant feature related to the NESS on which is worth dwelling. From this moment on, we only focus on the Liouvillian gap $\Delta_\lambda$, which is the non-vanishing eigenvalue of $\mathcal{L}$ with the smallest real part
\begin{equation}
    \Delta_\lambda \equiv -\underset{i}{\text{max}} \Re{\lambda_i}\,.
    \label{eq_def_liouvillian gap}
\end{equation}
This quantity controls the typical relaxation time of the longest-living eigenmode differing from the NESS. 

\subsection{Liouvillian gap at fixed $b$}
\label{sec_liouville gap fixed b}

\begin{figure}[!h]
    \centering
    \includegraphics[width=0.95\columnwidth]{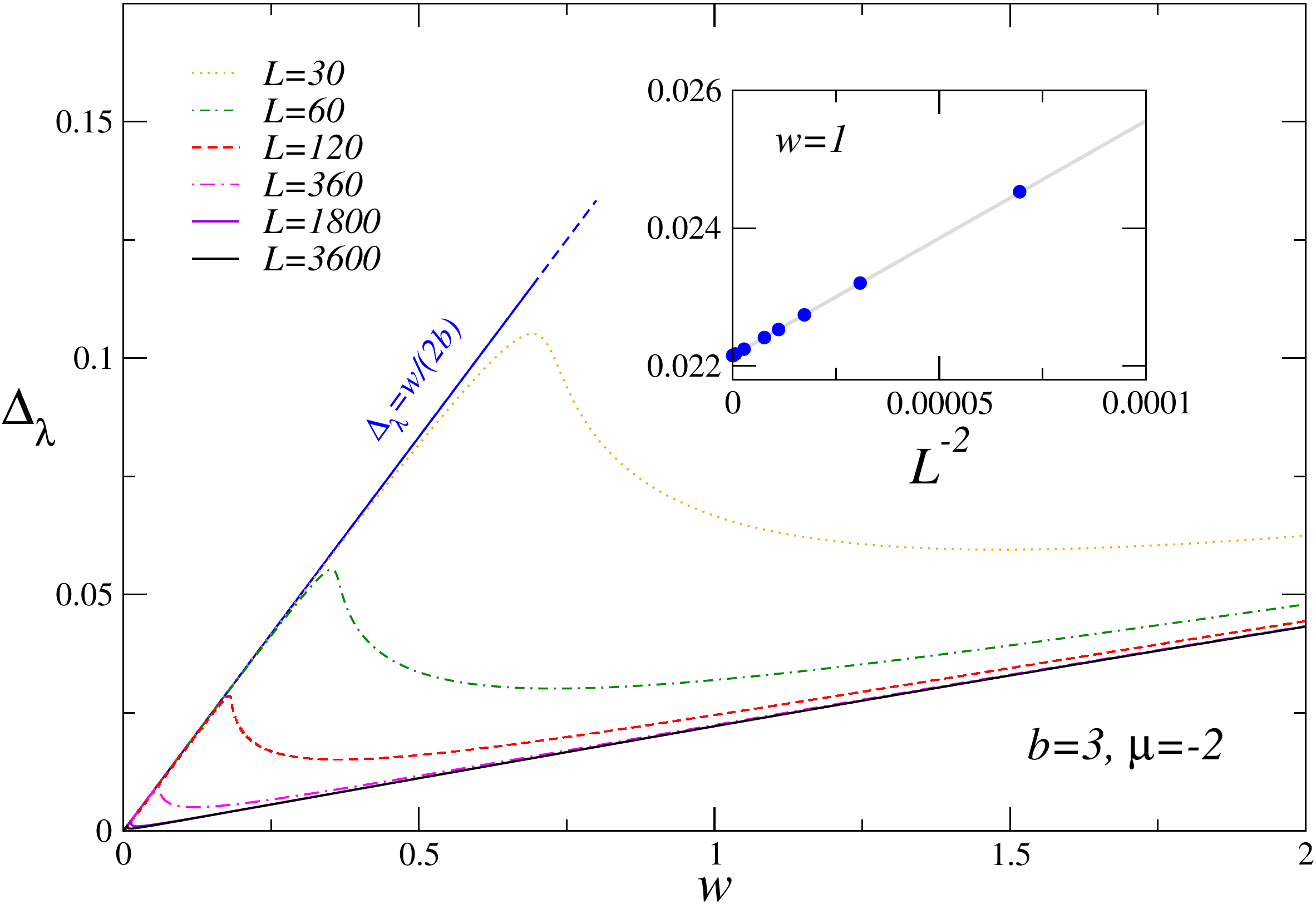}
    \caption{Liouvillian gap $\Delta_\lambda$ in terms of the dissipation coupling $w$ for $b=3$ and fixed $\mu=-2$. For small $w$ and $L$ finite, the gap depends linearly on the dissipation strength as $\Delta_\lambda=w/2b$. With increasing $L$ and finite $w>0$, the Liouvillian gap approaches a different regime, which still depends linearly on $w$. In the inset, scaling corrections evaluated at $w=1$ are perfectly consistent with a $L^{-2}$ decaying. The gray straight line is drawn to guide the eye.}
    \label{fig_liouvgap3b}
\end{figure}

We start our analyses by inspecting $\Delta_\lambda$ in the large-size limit with fixed $b$. For $b=1$, we prove that the model is always gapped for any $w>0$, see App.~\ref{sec_steadystate}. In particular, we show that independently of the chemical potential $\mu$ considered, the gap is equal to
\begin{equation}
    \Delta_\lambda=\frac{w}{2}\,.
    \label{eq_gap_b=1}
\end{equation}
We mention that a similar result has been observed for XXZ spin chains under the influence of dissipation~\cite{YWHWD-2021-artificialnetweork}. For $b>1$, the evaluation of both the NESS and $\Delta_\lambda$ is more complicated, and we are not able to provide a closed-form solution. However, since the Lindblad equations are quadratic in $\hat{c}$ and $\hat{c}^\dagger$, the Liouvillian gap can be obtained from the diagonalization of a $4L\times4L$ non-hermitian matrix using third quantization techniques~\cite{P-2008-thirdquantization}. Actually, by fully exploiting the residual translation invariance of the model under shifts of $b$, we reduce the numerical effort required by the algorithm after switching to the momentum basis. After this transformation, the gap $\Delta_\lambda$ can be retrieved from the diagonalization of $n/2$ matrices with dimension $8b\times8b$ (we always consider $L$ as a multiple of $2b$ when we work in momentum space), see App.~\ref{sec_app_simulation} for technical details on the algorithm. This strategy allows us to compute $\Delta_\lambda$ for lattice sizes up to $L\sim 3000$ for moderate values of $b\sim 3$, as we will see shortly.  

In the first part of the analysis, we set $\mu=-2$ and consider several values of $w$ and $b\leq7$. For $b=3$, we show our results for the Liouville gap $\Delta_\lambda$ in terms of $w$ in Fig.~\ref{fig_liouvgap3b}.
At fixed $L$, we can easily distinguish two different regimes for the gap, which are separated by a bump in the gap located at $w_*(L)$. We clarify that a non-monotone trend in $\Delta_\lambda$ is not unexpected due to the presence of the \textit{quantum Zeno effect} — the dynamic of a quantum system slows down when it is frequently monitored~\cite{MS-1977-quantumzeno, HHNU-2022-Incoherentons}. Note also that both $w_*(L)$ and $\Delta_\lambda(w_*)$ vanish in the thermodynamic limit, so only the region with $w\geq w_*$ is relevant to determine the typical relaxation time of the system for large enough ring sizes. As shown in Fig.~\ref{fig_liouvgap3b}, for $w<w_*$, the gap is perfectly compatible with a linear dependence of the form
\begin{equation}
\Delta_\lambda(w, b)=\frac{w}{2b}\,, \quad w<w_*\,.
\label{eq_deltaL_w_smaller_w*}
\end{equation}

Note that the equation comprises also the limiting case $b=1$, from which we get back Eq.~\eqref{eq_gap_b=1}. We have verified numerically that the above expression holds also for different values of $b\leq7$ (not shown). This equation has a clear interpretation when we rewrite $\mathbb{D}[\rho]$ in momentum space. Indeed, the full Hilbert space decomposes into the direct product of $n/2$ distinguished sectors with a dimension $4^b$. As derived in Eq.~\eqref{eq_momentum_space_dissipation_1_over_b} of App.~\ref{sec_app_simulation}, the effective coupling perceived within each sector is equal to $w/b$. If we additionally assume that the minimum contribution stemming from a single sector is $1/2$ (which is always the case for $b=1$), we get Eq.~\eqref{eq_deltaL_w_smaller_w*}.

\begin{figure}
    \centering
    \includegraphics[width=0.95\columnwidth]{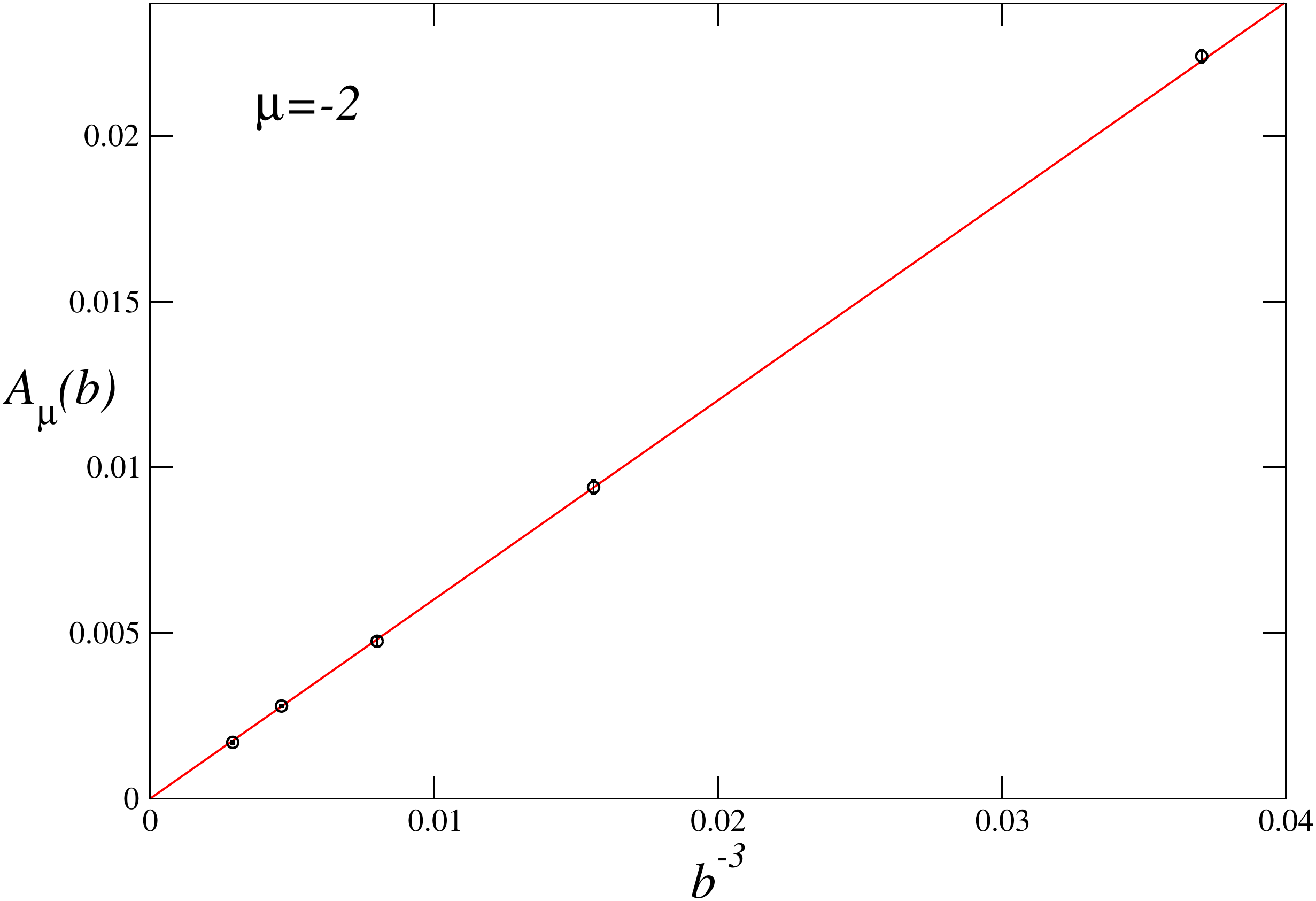}
    \caption{Liouvillian rate coefficient $A_{\mu}(b)$ versus $b^{-3}$ with constant $\mu=-2$. For $b\geq3$, we observe that $A_{\mu}(b)$ is compatible with a power-law dependence of the form $A_{\mu}(b)=C_{\mu}/b^{3}$, where $C_{\mu}=0.601(3)$ ($\chi^2/\text{ndof}=1.0$).}
    \label{fig_ratelatetime}
\end{figure}

On the other hand, when $w>w_*$, we observe that the gap $\Delta_\lambda$ still depends linearly on the coupling $w$, but the slope of the asymptotic straight line approached is no longer $1/2b$. We conjecture that for $w>w_*$ and sufficiently large $b$, the following expression describes the Liouvillian gap
\begin{equation}
    \Delta_\lambda(w, b)=A_\mu(b) w\,,\quad A_\mu(b)=\frac{C_\mu}{b^{3}}\,,\quad w>w_*\,,
    \label{eq_liouvillain_gap_largeL}
\end{equation}
where $C_\mu$ is a constant that only depends on the chemical potential $\mu$. Matching arguments with the boundary-dissipation cases surely prompted our guess. Indeed, when $b\propto L$, we expect to recover the leading behavior $\Delta_\lambda\sim L^{-3}$ frequently observed in the literature. Our ansatz is fully supported by the data that we have collected for $A_\mu$ in terms of $1/b^{3}$, as shown in Fig.~\ref{fig_ratelatetime} \footnote{Systematic error bars reported in the figure have been estimated from the comparison of $A_\mu(b)$ for different lattice sizes $L\geq L_{\text{min}}$ and coupling ranges $w\geq w_{\text{min}}$.}. Indeed, a straight line with a slope of $C_\mu=0.601(3)$ describes our data for all values of $b\geq3$ considered ($\chi^2/\text{ndof}=1.0$).  

\begin{figure}
    \centering
    \includegraphics[width=0.95\columnwidth]{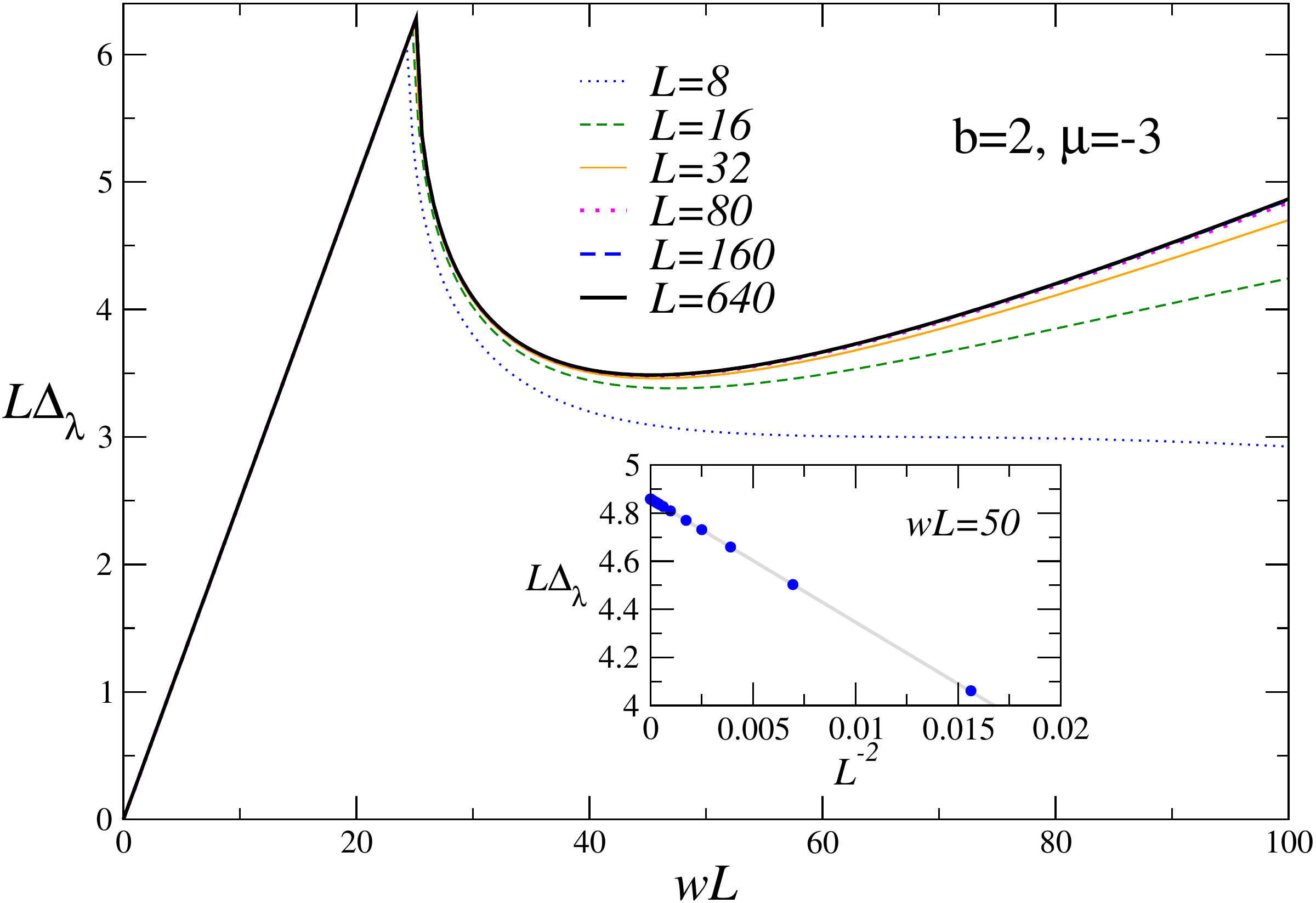}
    \includegraphics[width=0.95\columnwidth]{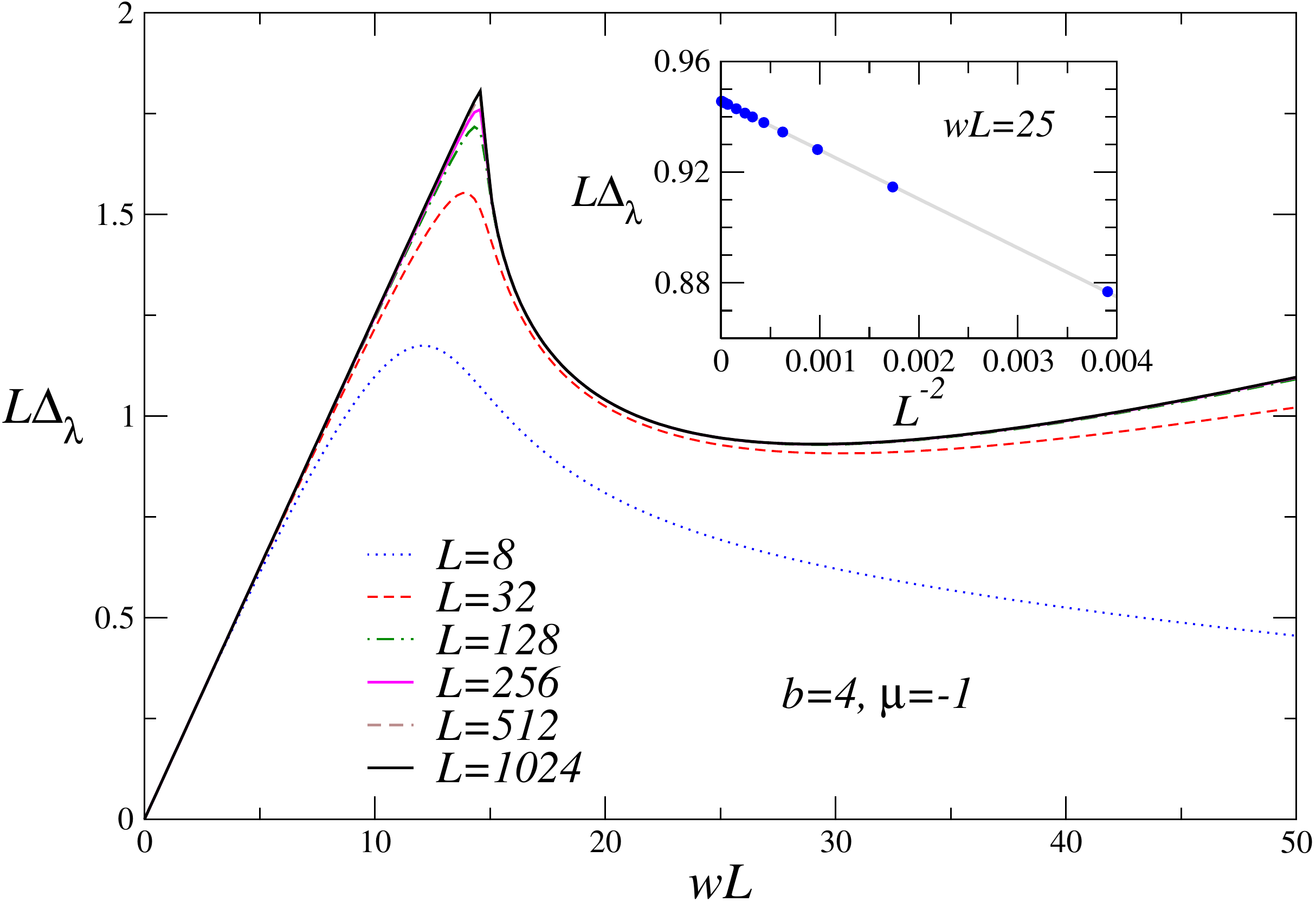}
    \caption{Scaling of $L \Delta_\lambda$ versus $w L$ for different values of $b$ and $\mu$. On the top panel, we show results for $b=2$ and $\mu=-3$, while on the bottom panel, we fix $b=4$ and $\mu=-1$. The figures show an excellent data collapse in agreement with $L^{-2}$ scaling corrections. The straight lines in the insets are drawn to guide the eye.}
    \label{fig_scaling_delta_apbc}
\end{figure}

We also want to mention a scaling regime observed for $L\Delta_\lambda$ in terms of $w L$ when the latter quantity is kept fixed in the large size limit. In Fig.~\ref{fig_scaling_delta_apbc}, we report our data for $b=2, \mu=-3$ and $b=4, \mu=-1$. In both panels, the data collapse that we observe is excellent along the whole curve. Of course, this scaling regime can be obtained from Eq.~\eqref{eq_deltaL_w_smaller_w*} and \eqref{eq_liouvillain_gap_largeL} for sufficiently small or large values of $w L$. However, the scaling hypothesis we propose works also in the intermediate regime $wL\sim w_*L$, capturing the behavior of $L\Delta_\lambda$ also at the peak. Apparently, our data suggest that the derivative of $L\Delta_\lambda$ shows a non-analytic behavior for $wL=w_*L$. Since our results in Fig.~\ref{fig_scaling_delta_apbc} are far from the QCP associated with the Kitaev ring, we also conclude that these scaling regimes are not controlled by universality arguments related to the quantum critical point.

\subsection{Liouvillian gap at fixed $n$}
\label{sec_liouvillian_gap_fixed_n}

In the following section, we study the dependence of the Liouvillian gap $\Delta_\lambda$ on the coupling strength $w$ when the number of dissipators is kept fixed. We employ third quantization techniques and work in the coordinate space to implement our algorithms, check App.~\ref{sec_app_simulation} for additional details concerning this section. Within our numerical capabilities, we explore lattice sizes up to $L\sim300$.

\begin{figure}
    \centering
    \includegraphics[width=0.95\columnwidth]{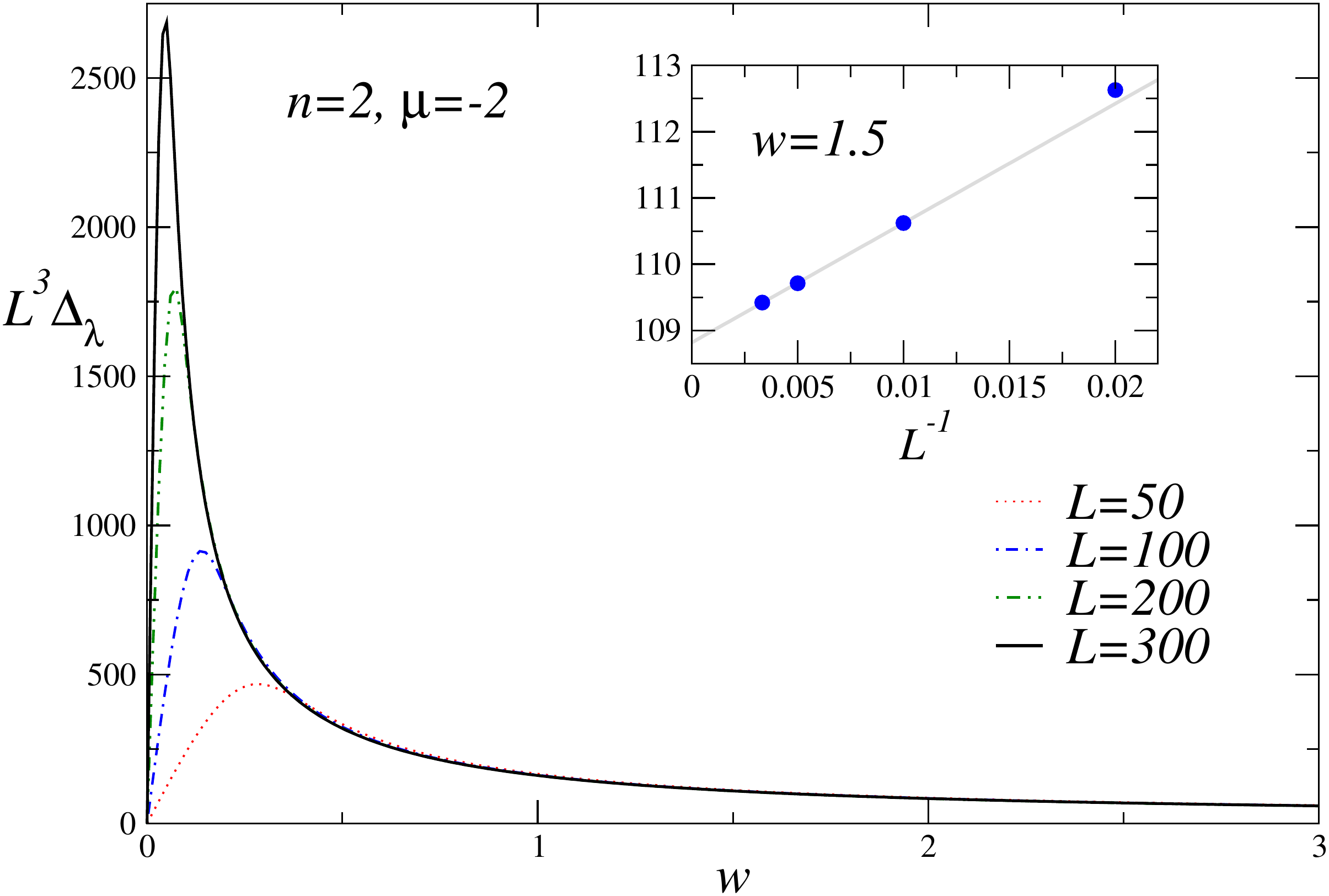}
    \caption{Plot of the rescaled gap $L^3\Delta_\lambda$ in terms of $w$ for fixed $n=2$ and $\mu=-2$.  At fixed $w$, the gap shows a nice data collapse within $L^{-1}$ scaling corrections, as provided by the inset for the case $w=1.5$. The straight line is drawn to guide the eye.}
    \label{fig_rescaleddelta_apbc_nfixed}
\end{figure}

First, we provide evidence of a gap vanishing as $\Delta_\lambda\sim L^{-3}$ at fixed $w$. This regime is supported by Fig.~\ref{fig_rescaleddelta_apbc_nfixed}, which shows a Kitaev ring with $n=2$ dissipators and $\mu=-2$. Indeed, for $w>w_*$, curves for different lattice sizes show a nice data collapse when we rescale the gap as $L^{3}\Delta_\lambda$. Scaling corrections are also compatible with a $L^{-1}$ decay, see the corresponding inset. This result is ascribed to Eq.~\eqref{eq_liouvillain_gap_largeL} and simple matching arguments. Indeed, since the distance among consecutive Lindblad operators increases linearly with $L$, that formula entails a scaling of the type $\Delta_\lambda\sim L^{-3}$. Note also that $\Delta_\lambda$ vanishes for all $w>0$, unlike the results of the previous section. 

\begin{figure}
    \centering
    \includegraphics[width=0.95\columnwidth]{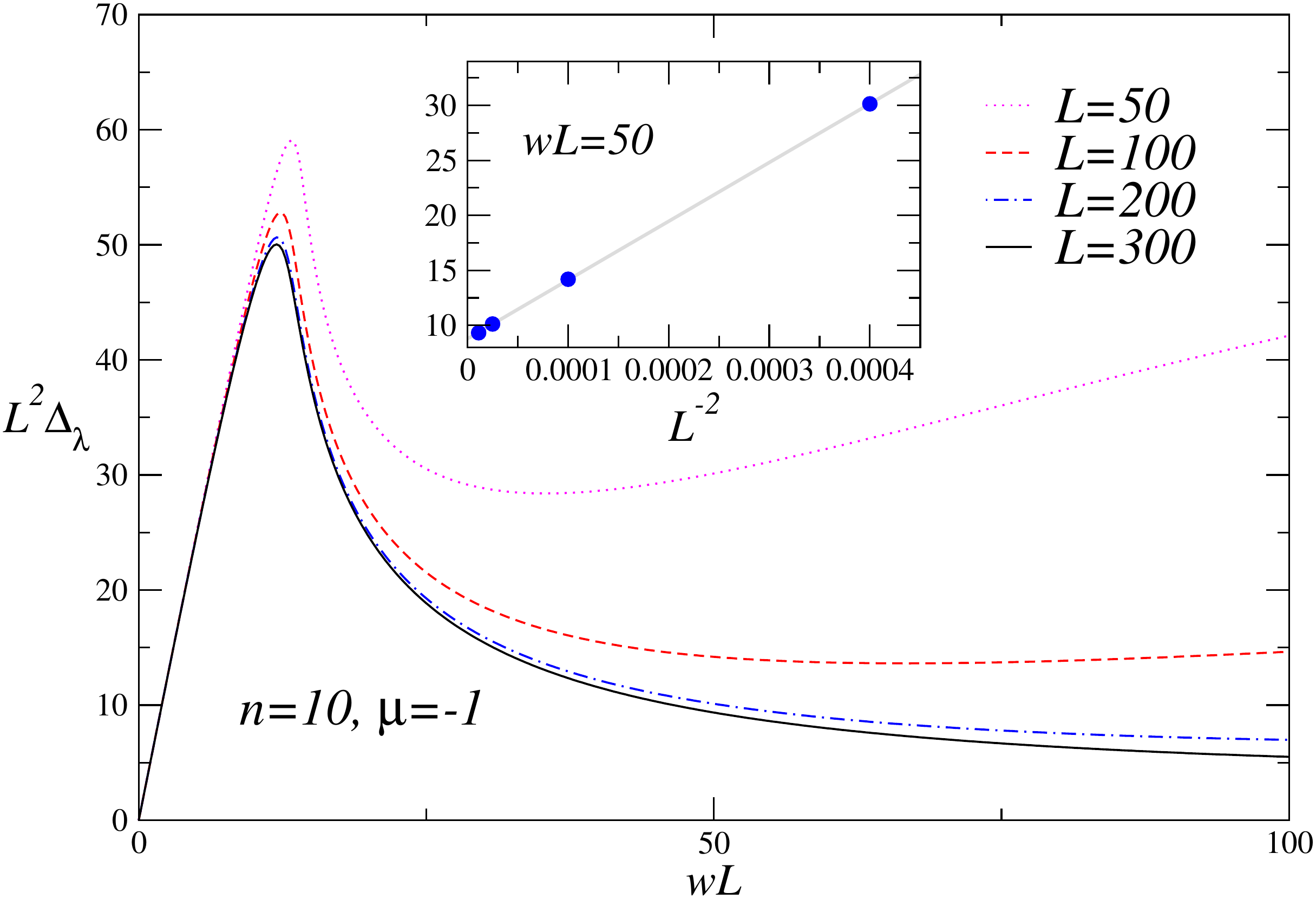}
    \caption{The figure shows the Liouvillian gap $L^2\Delta_\lambda$ versus $w$ at fixed $n=10$ for $\mu=-1$. In the inset, scaling corrections at $wL=50$ are consistent with a decay $L^{-2}$. The straight line is drawn to guide the eye.}
    \label{fig_scalingDeltanfixed}
\end{figure}

Referring to Fig.~\ref{fig_rescaleddelta_apbc_nfixed}, when $w<w_*$ the gap does not show a uniform limit for $w\to0^+$ as the maximum of $L^3\Delta_\lambda$ grows without bounds with increasing $L$. We shed some light on this peculiar trend in Fig.~\ref{fig_scalingDeltanfixed}, considering the structure of the gap in the proximity of $w=0$ at fixed $n=10$ and $\mu=-1$. In fact, the plot supports the existence of a scaling regime for $L^2\Delta_\lambda$ when $w$ is properly rescaled as $w \sim 1/L$. Scaling corrections are also compatible with a decaying $L^{-2}$, as shown in the corresponding inset. We stress again that numerical results for different values of $\mu$ do not exhibit remarkable differences. We conclude that the different scaling regimes shown by the Liouvillian gap do not depend either on $\mu$ or the quantum phase related to the Kitaev model.

\begin{figure}
\centering
\begin{tikzpicture}[scale=0.6]
    \draw[thick] (-6, 0) -- (7, 0);

    \foreach \x in {0,0.75,...,13}
    \filldraw [fill=blue!50] (-6+\x, 0) circle (0.2);

    \foreach \x in {0,3,...,13}
    \draw[orange, dashed, ultra thick] (-6+\x, 0.2) .. controls (-6+0.3+\x, 0.5) and (-6-0.3+\x, 0.8) .. (-6+\x, 1.1);

    \foreach \x in {0,3,...,12}
    \node[rectangle,
    draw = red,
    text = black,
    fill = red!20!white,
    rounded corners,
    minimum size=0.6cm] (r) at (-6+\x, 1.5) {\Large $\mathcal{B}$};

\end{tikzpicture}
\caption{Sketch of the Kitaev chain with OBC. This figure represents the case with $b=4$ and $L=18$ (the number of baths here is $n=5$).}
\label{fig_sketch_obc}
\includegraphics[width=0.95\columnwidth]{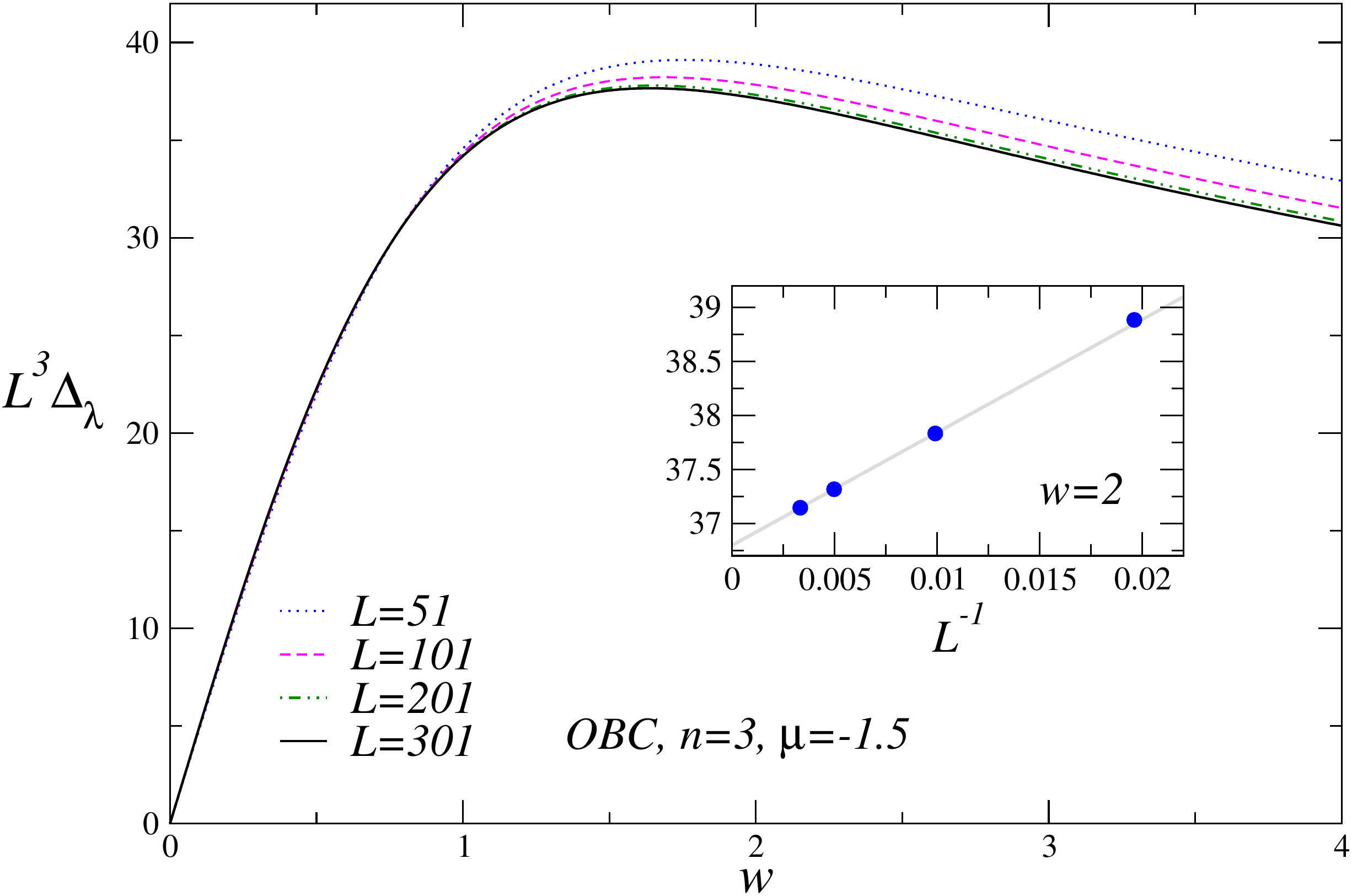}
\caption{The rescaled gap $L^3\Delta_\lambda$ in terms of the coupling $w$ for $\mu=-1.5$ and $n=3$. For all values of $w\leq4$ considered the rate of convergence appears uniform along the whole curves, differing from the case with APBC, cf. Fig.~\ref{fig_rescaleddelta_apbc_nfixed}.}
\label{fig_gap_bl_open}
\end{figure}

For the sake of completeness, we have also compared our results with the ones we would obtain using Open Boundary Conditions (OBC), considering a quantum open chain instead of a closed ring. A representative sketch of the lattice setup is reported in Fig.~\ref{fig_sketch_obc}. Notice that we always put a dissipator on the first site $x=1$, as given by Eq.~\eqref{eq_def_dissipator}. According to our conventions, the chain exhibits an invariance under spatial reflections (which sends $x\to L+1-x$) if and only if $L=kb+1$, with $k\in\mathbb{N}$. 
In Fig.~\ref{fig_gap_bl_open}, we present our results for $L^{3}\Delta_\lambda$ versus $w$ with OBC for $\mu=-1.5$ and $n=3$, considering two dissipators to the extremes and one at the center of the chain—we preserve spatial reflection symmetry. These data exhibit differences from the corresponding ones obtained in Fig.~\ref{fig_rescaleddelta_apbc_nfixed} with APBC. Even if both figures show a gap decaying as $L^{-3}$ for finite $w>0$, it is clear that a uniform convergence when $w\to0^+$ is only observed in the case of OBC. A similar mechanism was first appreciated in Ref.~\cite{TV-2022-localizedparticleloss}. In that paper, the authors addressed the study of the Liouvillian gap in a tight-binding model with OBC under the influence of a single particle-decay dissipator. They observed a non-uniform limit for $L^3\Delta_\lambda$ only when the dissipator was located at the center of the chain. As a result, this choice left all the odd modes untouched throughout time evolution. We suspect that a similar mechanism takes place in the Kitaev ring under study, generating the non-uniform behavior observed for $w\to0^+$. This issue requires further investigations to be better clarified.

\section{Dynamic Finite-Size Scaling (FSS) framework at CQT}
\label{sec_fss}

In this section, we study the time evolution of the Kitaev ring in the proximity of a CQT. To this end, we exploit a dynamic FSS framework and use RG arguments to describe the evolution of the critical correlations and the entanglement entropy. Concerning the algorithms adopted, we speed up our simulations by moving to the momentum basis every time we maintain $b$ fixed. This strategy allows us to explore lattice sizes up to $L\sim1500$ for relatively small $b\leq3$.  On the other hand, when $n$ is fixed, we just monitor the evolution of the two-point correlation functions by solving a closed system of differential equations. To evolve the density matrix $\rho$ in time, we use standard $4^{\text{th}}$-order Runge-Kutta techniques with typical integration time steps of $\Delta t=0.01$. All further details regarding the algorithms are postponed to App.~\ref{sec_app_coordinate_space}

\subsection{The quench protocol and the monitored observables}
\label{sec_observables}

We now present the quench protocol considered to study the time evolution of the open quantum system under scrutiny at CQTs. We prepare the system in the ground state $\ket{\Omega}$ of Eq.~\eqref{eq_kitaev}, so the density matrix is initially described by a \textit{pure state} given by $\rho=\ket{\Omega}\bra{\Omega}$. The starting chemical potential $\mu_i$ is always close to the critical value $\mu_c$, meaning that $\abs{\mu_i-\mu_c}\to0$ for $L\to\infty$. At a reference time $t=0$, the ring is driven out-of-equilibrium by suddenly coupling the system with the surrounding environment and eventually quenching the chemical potential to a different value $\mu_i\to\mu_f$. In such a case, the final $\mu_f$ should always be sufficiently close to the QCP. 

We monitor the time evolution of the Kitaev ring by considering two distinguished two-point correlation functions $C(x, y, t)$ and $P(x, y, t)$, defined as
\begin{align}
    C(x, y, t)&\equiv\Tr[\rho(t)(\hat{c}^\dagger_x\hat{c}_y+\hat{c}^\dagger_y\hat{c}_x)]\,,\\
    P(x, y, t)&\equiv\Tr[\rho(t)(\hat{c}^\dagger_x\hat{c}^\dagger_y+\hat{c}_y\hat{c}_x)]\,.
    \label{eq_def_two_point_functions_C_P}
\end{align}
To better characterize the dissipative processes, we also consider the Von-Neumann entropy $S(t)$ associated with the density matrix $\rho$ of the whole ring as
\begin{equation}
    S(t)=-\Tr[\rho(t)\log{\rho(t)}]\,.
    \label{eq_def_entanglement}
\end{equation}
The entropy just defined ranges from $0$, in the case of a pure state, to $L\log2\approx 0.69315.. \times L$, in the case of a maximally entangled state.


\subsection{Out-of-equilibrium FSS frameworks at CQTs with $b$ fixed}
\label{sec_Out-of-equilibrium FSS at CQT}

To describe the time evolution of the system under study at CQTs, we employ RG arguments and a dynamic FSS framework~\cite{C-1996-ScalingandRG, RV-2021-coherentanddissipativedynamicsreview}. The interplay between the unitary and dissipative dynamics of a Kitaev ring subject to complete bulk dissipation ($b=1$) has already been addressed in Ref.~\cite{NRV-2019-competingdissipativeandcoherent}. The results presented in this section extend the FSS reported in that work to all the cases with fixed $b>1$, and provide a complementing discussion on the role of $\Delta_\lambda$ in such a regime. Let us first review the main ideas leading to the FSS theory that we are going to discuss.

Sufficiently close to a continuous transition, the equilibrium and out-of-equilibrium scaling properties of a system are controlled by universality arguments associated with the nearby RG fixed point. To observe universal critical behaviors in the limit $L\to\infty$, all the parameters corresponding to relevant perturbations should be rescaled according to their RG scaling dimension~\cite{RV-2021-coherentanddissipativedynamicsreview}. To begin with, we introduce the scaling variable $M$ associated with the chemical potential $\mu$ as
\begin{equation}
    M=(\mu-\mu_c)L^{y_\mu}\,, \quad y_\mu=1\,.
    \label{eq_mu_scaling}
\end{equation}
This is the unique relevant perturbation to be taken into account to study the equilibrium properties of the Kitaev ring in the FSS limit. For instance, a generic observables $O(\{x_i\})$ with scaling dimension $y_O$, obeying standard FSS relations, satisfies
\begin{equation}
    O(\mu, L, \{x_i\}) \approx L^{-y_\mathcal{O}}\mathcal{O}(M, \{X_i\})\,.
    \label{eq_mathcalO_scaling}
\end{equation}
where $X_i\equiv x_i/L$, and $\mathcal{O}$ is a universal scaling function that only depends on the universality class related to the critical point and boundary conditions of the lattice.

When we consider the time evolution of an open quantum system after a quench, analogous equations are more involved given the presence of a larger number of scaling quantities and relevant perturbations. First of all, we need to introduce a pre- and a post-quench scaling field $M_{i/f}=(\mu_{i/f}-\mu_c)L^{y_\mu}$ for $\mu$. In the second place, the time variable $t$ requires a scaling field as well. The most natural guess, which also turns out to be the correct one in most cases, is to rescale $t$ with $L$ on the basis of the dynamic critical exponent $z$. We then introduce the quantity $\Theta$ defined as
\begin{equation}
    \Theta = t L^{-z}\,, \quad z=1\,,
    \label{eq_def_Theta_scaling}
\end{equation}
which is maintained constant in the FSS limit.
Since the number of particle-decay jump operators increases as $L$, we also need to soften the coupling $w$ to observe an interplay between the critical and dissipative modes. We note that the parameter $w$ plays the role of
a decay rate, namely, it is an inverse relaxation time~\cite{NRV-2019-competingdissipativeandcoherent, BP-openquantumsystembook}.
In our work hypothesis, we then suppose that $w$ should be rescaled with $L^{-z}$ to observe universal FSS relations. We introduce the scaling field $\gamma_b$ as
\begin{equation}
    \gamma_b=\frac{wL^{z}}{b}\,,\quad z=1.
    \label{def_gamma_scaling}
\end{equation}
Naturally, the prefactor $b^{-1}$ appearing in $\gamma_b$ is just a matter of convention if one restricts the analysis to just one value of $b$. However, the comparison between different values of $b$ in the FSS limit may add new valuable insights to our analyses. To compare the dissipative processes of different rings on the same footing, we assume that the effective coupling strength is $w/b$. This choice is the most natural one considering the Kitaev ring in momentum space. 

\begin{figure}
    \centering
    \includegraphics[width=0.95\columnwidth]{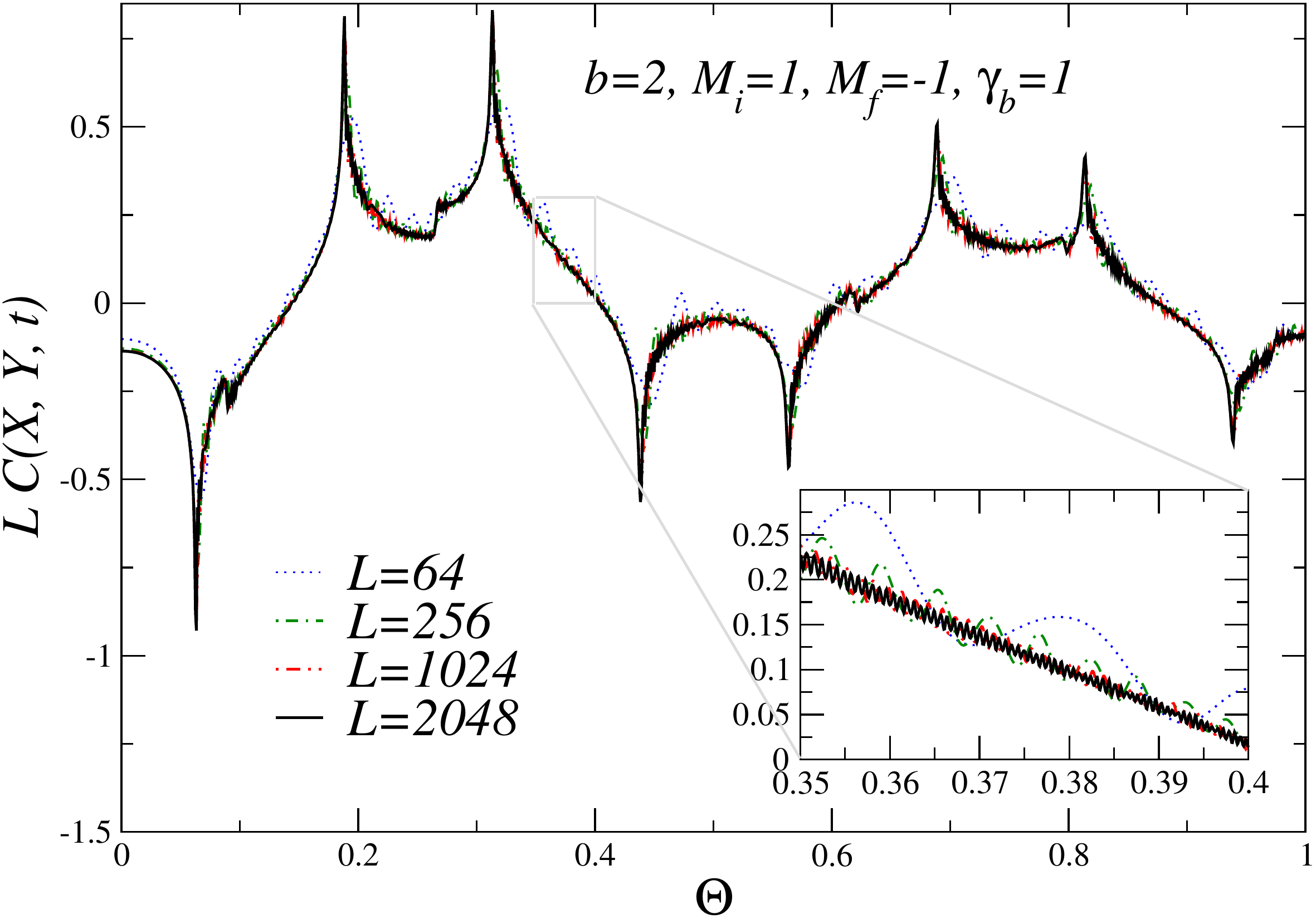}
    \caption{Scaling of the two-point function $L C(X, Y)$ in terms of the scaling variable $\Theta$ for fixed $b=2$, $M_i=1$ and $\gamma_b=1$. We consider $(Y-X)/L=1/4$ and fix $x=1$, using translation invariance. In the inset, we show a zoom of the region $\Theta\in[0.35, 0.4]$. Our data clearly support the FSS laws exhibited in Eq.\eqref{eq_scaling_C}.}
    \label{fig_Cscaling2b}
\end{figure}

The universal scaling relations satisfied by the two-point functions $C$ and $P$ in Eq.~\eqref{eq_def_two_point_functions_C_P} follows
\begin{align}
    C(x, y, t) &\approx L^{-2y_c}\mathcal{C}(M_i, M_f, \{X_i\}, \Theta, \gamma_b)    \label{eq_scaling_C}\\
    P(x, y, t) &\approx L^{-2y_c}\mathcal{P}(M_i, M_f, \{X_i\}, \Theta, \gamma_b)\,,
    \label{eq_scaling_P}
\end{align}
where $y_c=1/2$ is the scaling dimension of both $\hat{c}$ and $\hat{c}^\dagger$. 
In Fig.~\ref{fig_Cscaling2b} we show the scaling of $LC(X, Y, t)$ in terms of the scaling quantity $\Theta$ for $b=2$, $M_i=1$, $M_f=-1$, and $\gamma_b=1$. The panel definitely supports the FSS laws exhibited in Eq.~\eqref{eq_scaling_C}. In the inset, we show that the amplitude of the oscillations reduces at fixed $\Theta$ and increasing $L$, roughly as $\sim L^{-1/2}$. We have checked that Eq.~\eqref{eq_scaling_P} holds also for the scaling of the RG invariant quantity $LP(X, Y, t)$ (not shown).  

\begin{figure}
    \centering
    \includegraphics[width=0.95\columnwidth]{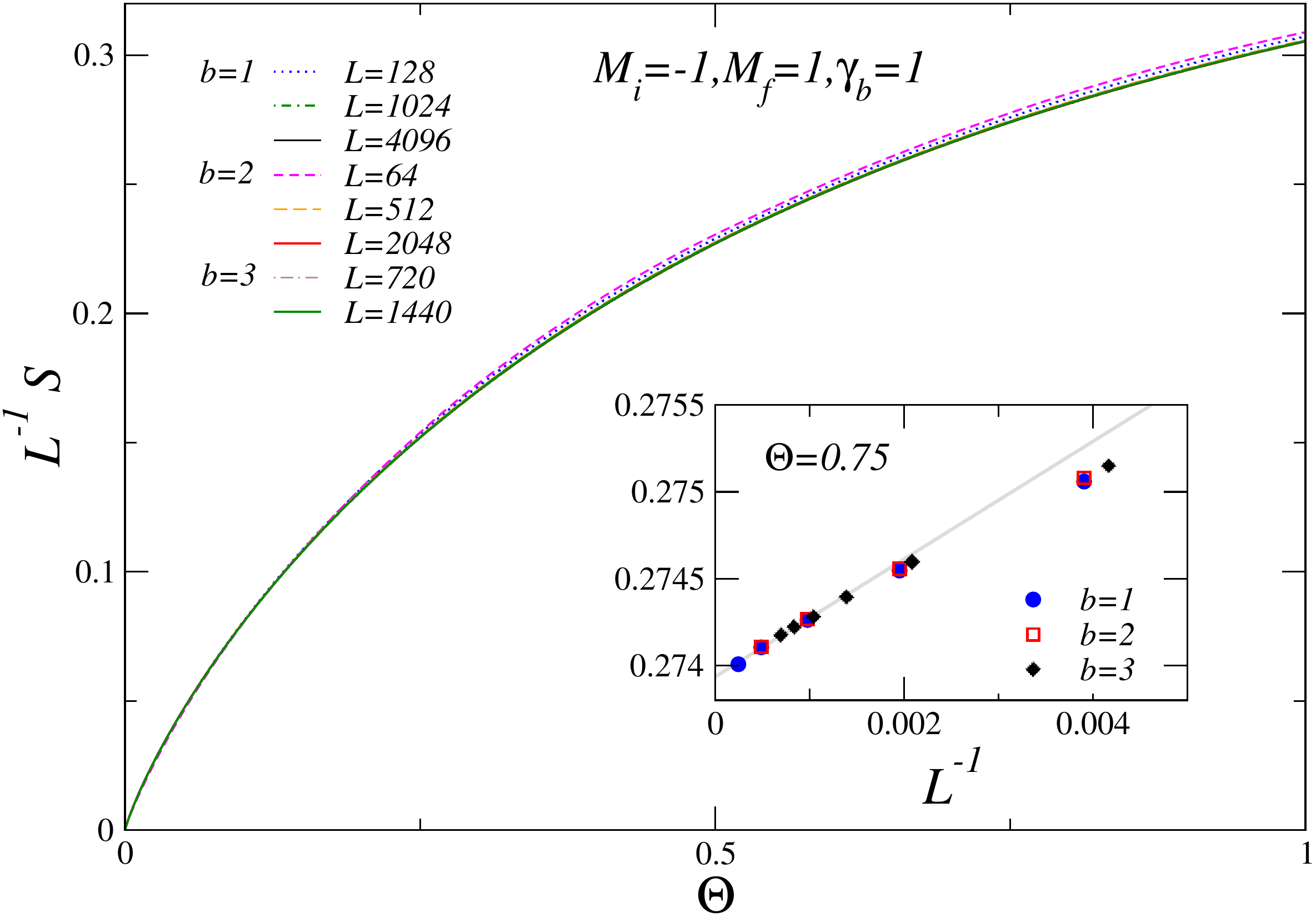}
    \caption{Scaling of the entanglement entropy per unit-length $S(t)/L$ in terms of the scaling variable $\Theta$ at fixed $M_i=-1, M_f=1$, and $\gamma_b=1$. In the inset, scaling corrections are consistent with a $L^{-1}$ decay at fixed $\Theta=0.75$. The gray straight line is drawn to guide the eye.}
    \label{fig_entanglementscaling}
\end{figure}

Successively, we address the feasibility of universal FSS relations unifying the critical behavior of open Kitaev rings with different $b$. We anticipate that some of the observables we have considered corroborate this hypothesis, but others do not. The most compelling evidence in favor of a universal scaling is provided by the entanglement entropy $S(t)$. Since this observable is an extensive quantity, we conjecture that the entanglement entropy per unit-length $S(t)/L$, cf. Eq.~\eqref{eq_def_entanglement}, is an RG invariant quantity at the QCP.
\begin{equation}
    S(t) \approx L \mathcal{S}(M_i, M_f, \Theta, \gamma_b)\,.
    \label{eq_entanglement_fss}
\end{equation}
We emphasize that this quantity and the entanglement entropy associated with disjoint intervals in $(1+1)$-dimensional \textit{closed} systems do not share the same critical behavior~\cite{CC-2009-entanglementcritical}. In fact, well-established results hold for the latter quantity at equilibrium, diverging as $\propto \log L$ at the critical point. 
We examine several values of $b\leq3$ and fix all the relevant scaling variables to $M_i=1$, $M_f=-1$, and $\gamma_b=1$. Our results are shown in Fig.~\ref{fig_entanglementscaling}. The data collapse we obtain is surprisingly good if we consider that the scaling corrections observed decay only as $\sim L^{-1}$. 

\begin{figure}
    \centering
    \includegraphics[width=0.95\columnwidth]{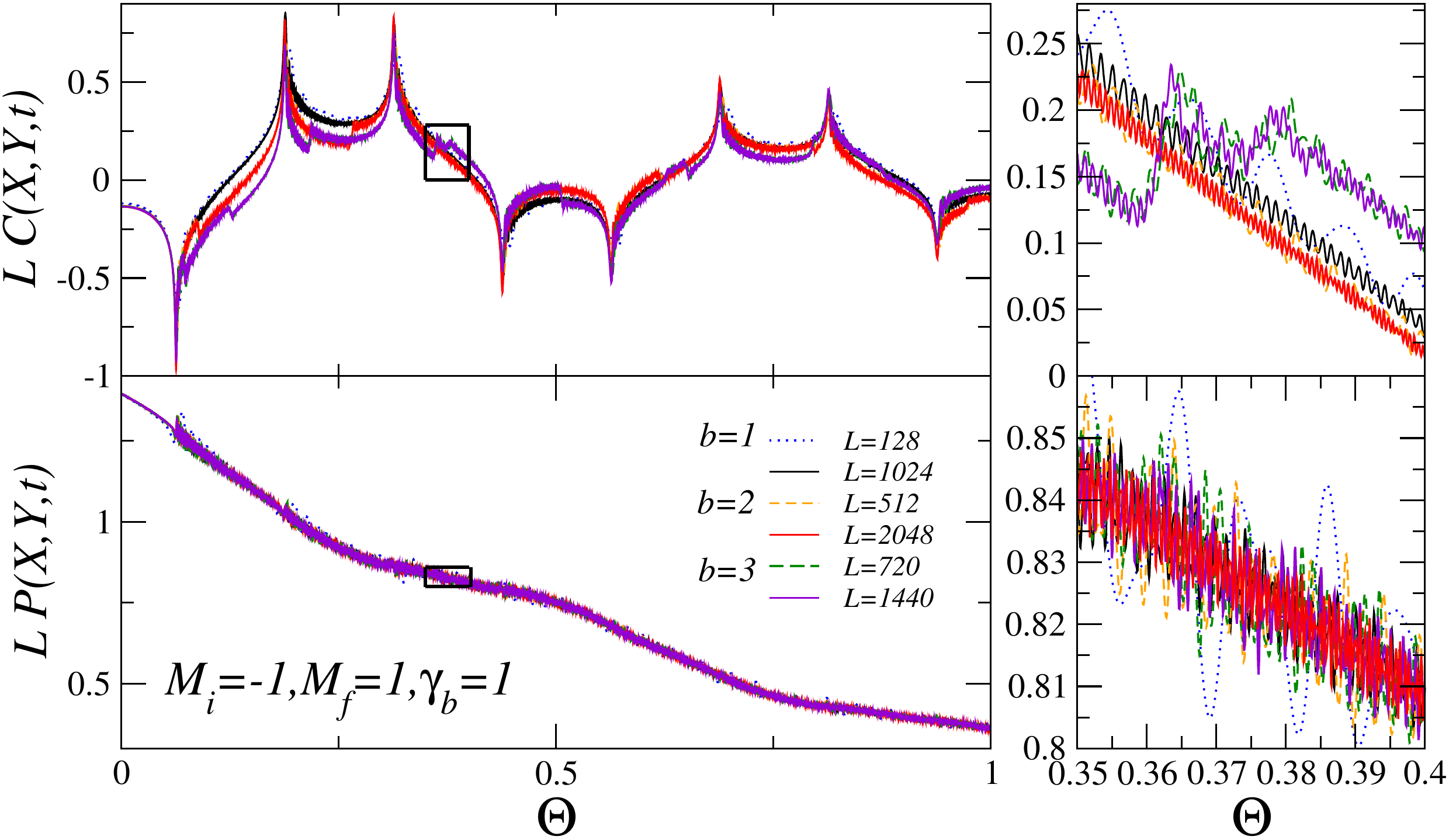}
    \caption{Scaling of two-point correlation functions $L C(X,Y,t)$ and $L P(X,Y,t)$ versus the scaling variable $\Theta$, respectively in the upper and bottom panel (again $Y-X=L/4$). In the insets, we zoom in on the domains which are boxed in leftmost plots, considering only the region with $\Theta\in[0.35, 0.4]$.}
    \label{fig_CPscalingmanyb}
\end{figure}

The same analysis is applied to the critical correlations $L C(X, Y, t)$ and $L P(X, Y, t)$ with the aim of verifying whether open rings with different $b$ are described by the same FSS relations. Our results are not completely clear on this point, see for example Fig.~\ref{fig_CPscalingmanyb}. The plot of $L C(X, Y, t)$, in the top panel, does not support a universal behavior for lattice models with different $b$. Even if the figure shows the same pattern of spikes in $L C(X, Y, t)$, entailing some sort of connection between the time evolution of the modes associated with different $b$, we cannot conclude that they all share the same universal scaling function $\mathcal{C}$. Zooming in on the region $\Theta\in[0.35, 0.4]$ we rule out this hypothesis for the $C(x, y, t)$ correlations, see the inset on the right of Fig.~\ref{fig_CPscalingmanyb}. We cannot come to the same conclusion observing the plot of $L P(X, Y, t)$, which instead shows a nice scaling for all $b$ we consider. With increasing $L$, the oscillation amplitudes are all shrinking and apparently converging to the same $\mathcal{P}$. We cannot exclude that the scaling laws we have put forward in Eq.~\eqref{eq_scaling_C} for $L P(X, Y, t)$ may hold irrespectively of $b$.

To get to the point, we cannot neglect the robust evidence provided by the scaling of the entanglement entropy and the two-point correlation function $LP(x, y, t)$, reported respectively in Fig.~\ref{fig_entanglementscaling} and ~\ref{fig_CPscalingmanyb}. These results lead us to conclude that the scaling field $\gamma_b$ is a good scaling quantity in the FSS limit, which allows us to compare the dissipative mechanisms of different rings at the same level. However, the critical two-point function $LC(x, y, t)$ does not satisfy a unique universal FSS relation for all $b$, as far as our simulations allow us to conclude.

\subsection{Out-of-equilibrium FSS framework at CQTs with $n$ fixed}
\label{sec_out-of-equilibrium FSS fixed n}

\begin{figure}
    \centering
    \includegraphics[width=0.95\columnwidth]{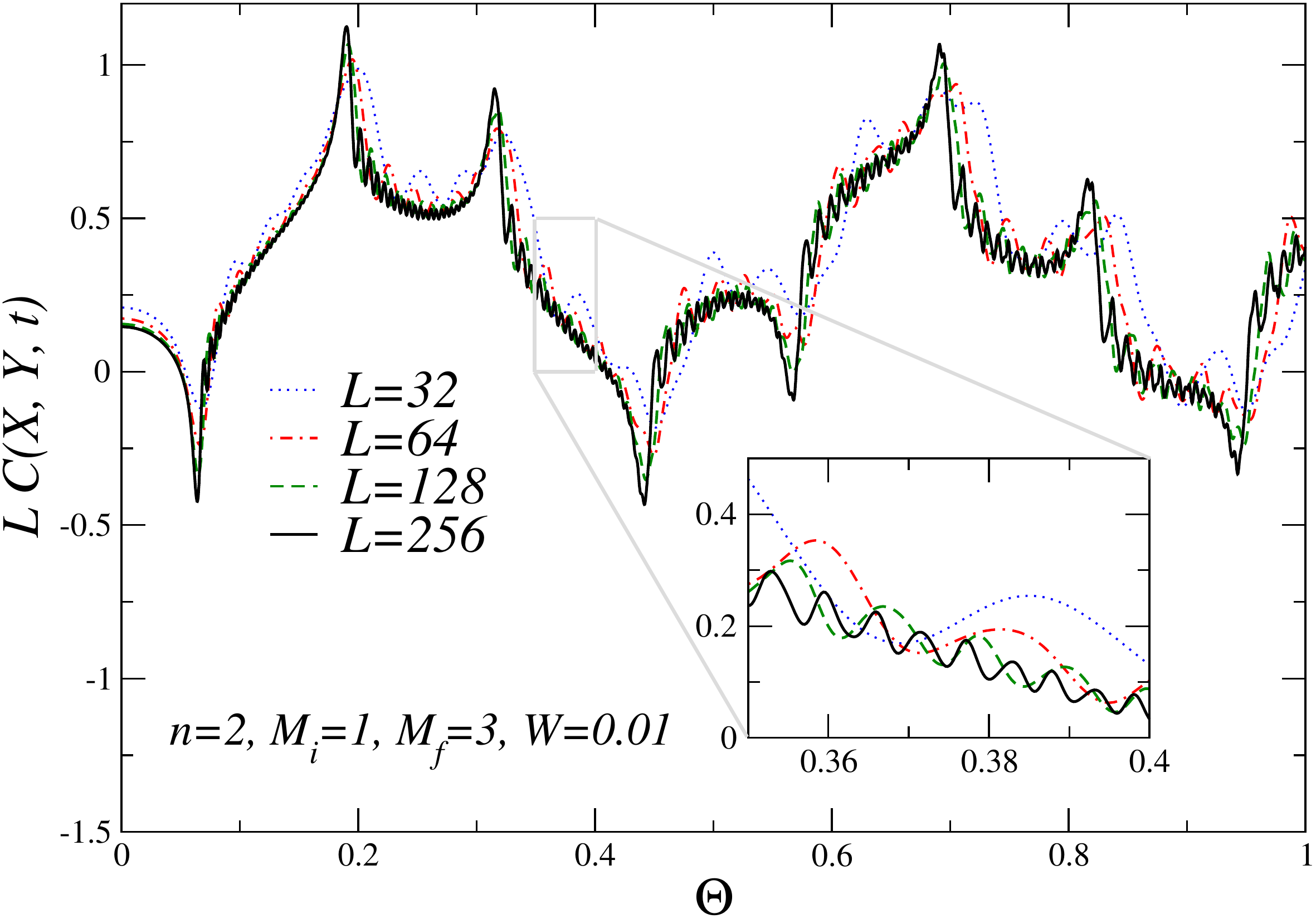}
    \caption{Scaling of the rescaled two-point correlation function $L C(X, Y)$ in terms of $\Theta$ for fixed of $n=2, M_i=1, M_f=3$, and $W=0.01$. We consider $Y-X=L/4$, exploiting translation invariance, and fix the first site to $x=1$ in conjunction with a dissipator. We zoom in on the domain $\Theta\in[0.35, 0.4]$ to emphasize the convergence of different curves with increasing lattice size $L$.}
    \label{fig_fss_2n_fixed}
\end{figure}

In this section, we derive a dynamic FSS framework at CQTs to study the time evolution of Eq.~\eqref{eq_def_liouvillian} when the number of the dissipators $n$ is fixed. Most of the scaling relations of the last section are still valid at fixed $n$ with straightforward generalizations. In particular, we just replace the coupling $w$ in our relations, which now represents a decay rate per unit space since $b\sim L$. As a working hypothesis, we expect the scaling field $W$, defined as
\begin{equation}
    W = w L^{z-1}\,, \quad z=1\,,
    \label{eq_def_scaling_variable_W}
\end{equation}
to be a reasonable scaling quantity in the FSS limit. 
For instance, the critical correlations satisfy scaling relations similar to the ones reported in Eq.~\eqref{eq_scaling_C}
\begin{align}
    C(x, y, t)\approx& L^{-2y_c}\mathcal{C}(M_i, M_f, \{X_i\}, \Theta, W)\\
    P(x, y, t)\approx& L^{-2y_c}\mathcal{P}
    (M_i, M_f, \{X_i\}, \Theta, W)\,.
    \label{eq_def_CP_scaling_n_fixed}
\end{align}

We verify our hypotheses in Fig.~\ref{fig_fss_2n_fixed}, showing the scaling curve for $L C(X, Y, t)$ versus the scaling variable $\Theta$ with constant $n=2, M_i=1, M_f=3, W=0.01$. We obtain a nice data collapse considering lattice sizes up to $L=256$. The oscillation amplitudes shrink with increasing $L$, converging to a universal asymptotic curve $\mathcal{C}$. The validity of Eq.~\eqref{eq_def_CP_scaling_n_fixed} has been checked also by inspecting the time evolution of $L P(X, Y, t)$ (not shown).

\subsection{The role of the Liouvillian gap $\Delta_\lambda$ in the FSS limit}
\label{sec_role_gap_in_FSS}

Up to this point, we have not mentioned either the Liouvillian gap $\Delta_\lambda$ or the asymptotic steady-state in the dynamic FSS theory put forward at CQTs. However, to maintain $\Theta=tL^{-z}$ fixed, we explore progressively longer times with increasing $L$. Therefore, we want to understand whether the early-$\Theta$ regime, which is controlled by universality arguments related to the QCP, connects smoothly with the large-$\Theta$ domain, which is instead controlled by the NESS. Note also that two distinguished time scales regulate the unitary and dissipative processes concerning these regimes. They are the gap associated with the CQT, i.e., $\Delta\sim L^{-z}$ with $z=1$, and the Liouvillian gap $\Delta_\lambda$. As regards the interplay between these two quantities, we first observe that a necessary condition to have a non-trivial FSS regime is given by 
\begin{equation}
    \lim_{L\to\infty}\frac{\Delta}{\Delta_\lambda}>0\,.
    \label{eq_necessary_condition_fss}
\end{equation}
In the opposite case, the NESS entirely controls the dynamic, and no universal relation is allowed. The above equation is always fulfilled in our analyses presented in Sec.~\ref{sec_fss}, either at fixed $b$ or $n$. 

To carry on our discussion, we introduce the RG invariant quantity $R_n$ defined as 
\begin{equation}
    R_n = \frac{N(t) - N_{\text{asy}}}{N(0) - N_{\text{asy}}}\,,
    \label{def_Rn}
\end{equation}
where $N(t)=\langle\hat{n}(t)\rangle$ and $N_{\text{asy}}=\lim_{t\to\infty}N(t)$. Unlike the two-point functions $C(x, y, t)$ and $P(x, y, t)$, this quantity does not present sharp high-frequency oscillations in the FSS limit and, for this reason, is more useful for the following discussion. 

\begin{figure}
    \centering
    \includegraphics[width=0.95\columnwidth]{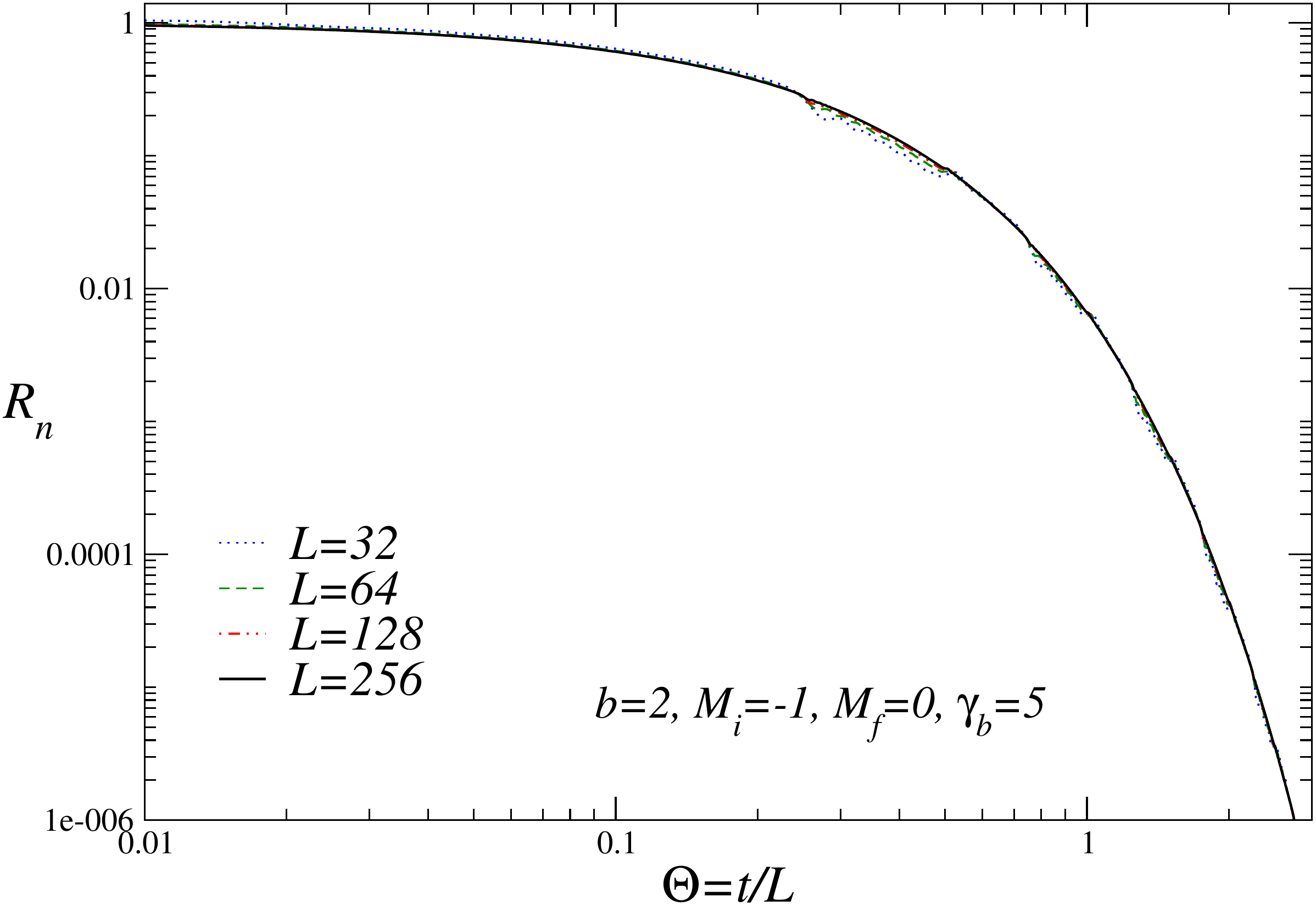}
    \caption{Scaling of the RG invariant quantity $R_n$ versus $\Theta=t/L$. This ratio decreases towards the NESS at larger times under the influence of dissipation. The data collapse we observe is excellent even at $\Theta\sim3$ where $R_n$ reduces from its starting value by a factor $\sim 10^{-6}$.}
    \label{fig_fss_delta_L_2b}
\end{figure}

We first address the roles of both $\Delta$ and $\Delta_\lambda$ in the FSS limit at fixed $b$. We maintain $\gamma_b=wL^zb^{-1}$ constant, so we expect both of the gaps to vanish with the same power-law behavior as $\sim L^{-1}$. We consider the RG invariant quantity $R_n$ in terms of $\Theta=t/L$ in Fig.~\ref{fig_fss_delta_L_2b} with $b=2$, $M_i=-1$, $M_f=0$, and $\gamma_b=5$. This quantity decreases to zero with increasing $\Theta$, showing an excellent data collapse along the whole curve. On the right of the figure, at $\Theta=3$, we can reasonably suppose that the data are sufficiently close to the asymptotic regime since the ratio $R_n$ significantly reduces its value by a factor of $10^{-6}$. The plot suggests that the early-$\Theta$ and large-$\Theta$ regimes connect smoothly within the finite-size scaling framework presented in this paper.  

\begin{figure}
    \centering
    \includegraphics[width=0.95\columnwidth]{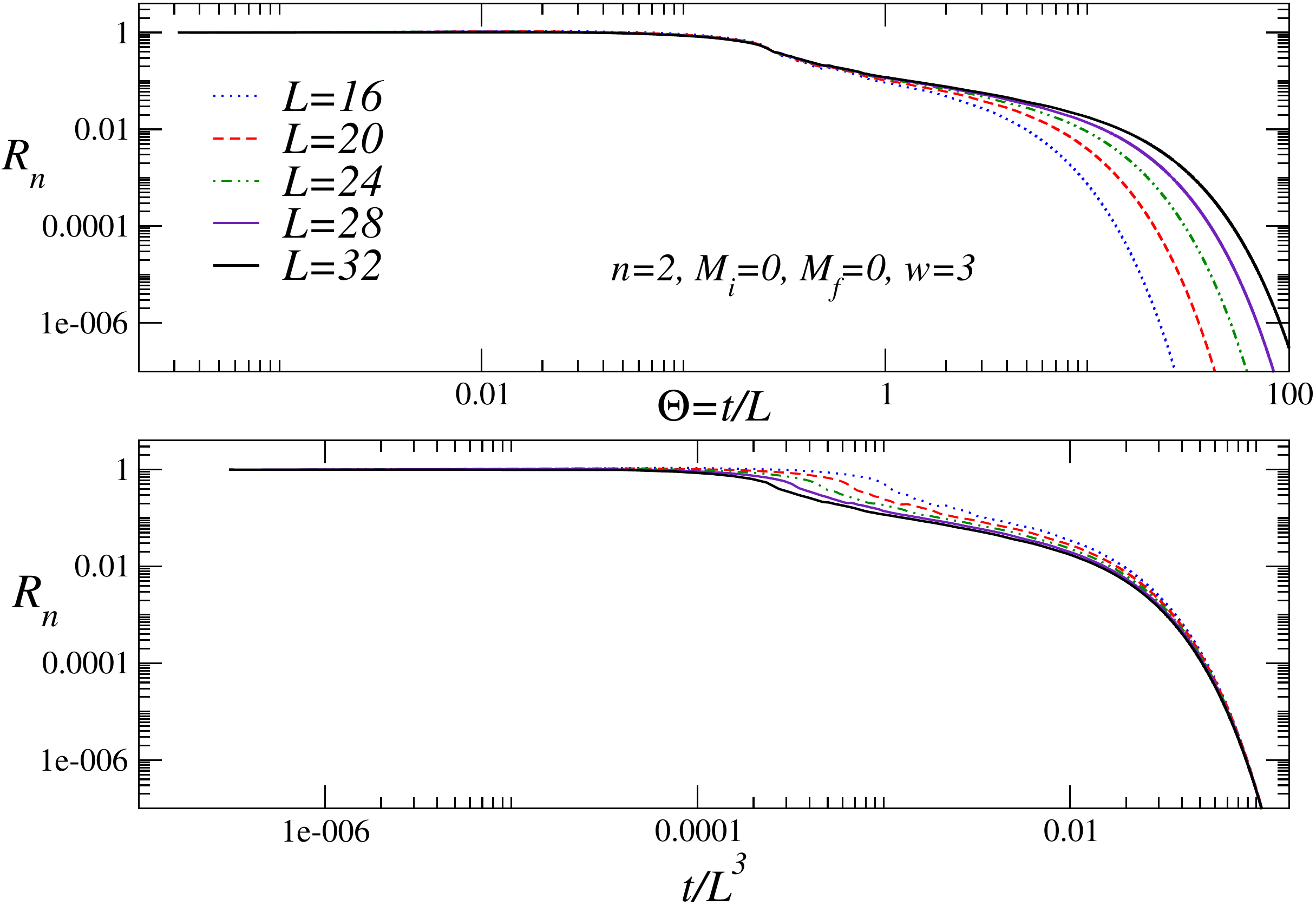}
    \caption{Top: The RG invariant quantity $R_n$ versus $\Theta$. The small-$\Theta$ behavior is controlled by the dynamic critical exponent $z$ according to the FSS theory put forward in Sec.~\ref{sec_fss}. Bottom: The large-time behavior of $R_n$ scales as $t/L^3$ at longer times since the Liouvillian gap $\Delta_\lambda$ controls the relaxation times of the model.}
    \label{fig_fss_gap_n_fixed}
\end{figure}

We now repeat the same analysis keeping the number of particle-decay dissipators fixed in the FSS limit. In this case, we get $\Delta_\lambda\sim L^{-3}\ll\Delta$ at the quantum critical point. We present our results in Fig.~\ref{fig_fss_gap_n_fixed} for $n=2$, $M_i=0$, $M_f=0$, and $w=3$ fixed. We only consider small lattice sizes (up to $L=32$) as we expect the asymptotic steady-state to emerge in a large amount of time of the order of $t\sim L^3$. In the upper panel, we rescale the time variable $t$ with the gap associated with the QCP, therefore considering $R_n$ versus $\Theta=tL^{-z}$. This RG invariant quantity shows a short-time regime controlled by $\Delta$ for small $\Theta$, but the data appear scattered at longer times. In fact, in the bottom panel of Fig.~\ref{fig_fss_gap_n_fixed}, we confirm that the Liouvillian gap controls the scaling regime of $R_n$ at large $t$—the rightmost data in the figure fall onto each other when we plot $R_n$ in terms of $t/L^3$. We conclude that by keeping $n$ fixed, the short- and large-time regimes of $R_n$ are not smoothly connected, being characterized by two physical mechanisms sharing different power-law scalings. 

\section{Conclusions}
\label{sec_conclusions}

In this paper, we have considered a $(1+1)$-dimensional Kitaev ring coupled with the environment via $n$ particle-decay dissipators arranged in a sunburst geometry.

In the first part of this work, we focus on the dependence of the Liouvillian gap $\Delta_\lambda$ on $L$ using different schemes to approach the large-size limits. When we keep $b$ fixed, the gap $\Delta_\lambda$ is always finite and depends linearly on the dissipation strength $w$. Nonetheless, two different regimes emerge for systems of finite size. In the small $w$ region, the gap is given by $\Delta_\lambda=w/(2b)$, whereas, at large $w$ and sufficiently large $b$, it behaves as $\Delta_\lambda=w C_\mu/b^3$. The last equation always controls the gap in the large-size limit and is our starting point to deduce the scaling of such a quantity when $b\propto L$. It is worth mentioning that we also put forward a scaling regime for $L\Delta_\lambda$ as a function of $wL$, which ties together the two different regimes outlined in a smooth manner. On the other hand, when we keep the number of dissipators $n$ fixed, the gap vanishes as $\sim L^{-3}$ at large $L$. Addressing the structure of the gap at small $w$, we find a scaling regime for $L^2\Delta_\lambda$ in terms of $wL$, which is closely related to the presence of a non-uniform convergence of $L^3\Delta_\lambda$ in the limit $w\to0^+$.

In the second part of this work, we develop a dynamic FSS regime at CQTs to describe the time evolution of the Kitaev model under investigation. At fixed $b$, our results extend the FSS theory of Ref.~\cite{NRV-2019-competingdissipativeandcoherent} to the cases with $b>1$. As a working hypothesis, we suppose that the scaling variable associated with the relevant coupling $w$ is $\gamma_b=wL^{z}/b$. Our numerical results for the two-point correlation functions and the entanglement entropy fully support this ansatz. In the second stage, we compare the real-time evolution of several rings corresponding to different $b$ to get some additional insights into the dissipation mechanisms of these systems.  As far as our numerical capabilities allow us to conclude, the entanglement entropy and the $P$-correlations admit a universal scaling function for all $b$, but the $C$-correlations do not. This issue requires further investigations to be better understood. When the number of dissipators $n$ is fixed, the FSS theory outlined at fixed $b$ generalizes straightforwardly after replacing $\gamma_b$ with $W=wL^{z-1}$. In the last section, we analyze the interplay between the Liouvillian gap $\Delta_\lambda$ and the gap related to the Kitaev ring $\Delta$ in the FSS limit. In particular, we take into account the short- and long-time regimes, focusing on how they join together in the FSS limit. When $b$ is fixed, we observe that the link between the two regimes is smooth, whereas, at fixed $n$, the two regions can be easily distinguished given the presence of different power-law scalings for the gaps $\Delta$ and $\Delta_\lambda$.  

As future outlooks, we mention that the results of this work can be extended in several directions. First of all, our studies can be generalized by considering thermal baths in the Lindblad formalism. Alternatively, it would be interesting to understand how the different large-size limits considered in this paper affect the Liouville gap and the FSS regime of open quantum models in higher dimensions. Despite the numerous challenges given by such a quest, we must say that this setting certainly offers attractive questions and new paradigms to be addressed. To name a few, we mention that the NESS, in more than one spatial dimension, may undergo a continuous phase transition, similar to a finite-temperature quantum system at equilibrium. For this reason, the evolutions of open quantum models in the short- and long-time regimes can be associated with different RG fixed points, entailing a more intriguing scenario in the FSS limit.

\section*{Acknowledgment}
The authors are delighted to thank E. Vicari for useful observations and stimulating discussions on this paper.

\appendix

\section{Steady-state solution and $\Delta_\lambda$ for $b=1$.}
\label{sec_steadystate}
In this appendix, we discuss the spectrum of the Liouville superoperator $\mathcal{L}[\rho]$ appearing in Eq.~\eqref{eq_def_liouvillian} for the case $b=1$. In particular, we focus on the Liouvillian gap $\Delta_\lambda$ and the steady-state solution. The study is dramatically simplified after we move to the momentum basis. To this end, let us first review the unitary Kitaev ring in momentum space in the absence of dissipation. 

We define the Fourier transform of the operator $\hat{c}_x$ as~\cite{NRV-2019-competingdissipativeandcoherent}
\begin{equation}
    \hat{c}_x=\frac{e^{-i\pi/4}}{\sqrt{L}}\sum_{k}e^{ikx}\hat{c}_k\,, \quad k=\bigg\{\pm\frac{(2n-1)\pi}{L}\bigg\}\,,
\end{equation}
where the momenta are induced by the boundary conditions used and $n=1,\dots, L/2$. To simplify the discussion, we only consider even lattice sizes, so that $L/2$ is always an integer number. For each mode $k>0$, we can choose an ordered Hilbert-space basis of the form $\{\ket{0_k},\ket{1_k},\ket{1_{-k}}, \ket{1_{k,-k}}\}$, where the hamiltonian is $\hat{H}=\sum_{k>0}\hat{H}_k$ with
\begin{equation}
    \hat{H}_k=\begin{pmatrix}
    0 & 0 & 0 & 2\abs{\sin k}\\
    0 & -2f_k(\mu) & 0 & 0\\
    0& 0 & -2f_k(\mu)&0\\
    2\abs{\sin k}& 0& 0& -4f_k(\mu)\\
    \end{pmatrix}\,,
    \label{eq_app_Hk_b1}
\end{equation}
and $f_k=\mu/2+\cos k$~\cite{NRV-2019-competingdissipativeandcoherent, P-2008-thirdquantization}. The full Hilbert space $\mathcal{H}$ decomposes naturally into the direct product of $n$ distinct $4$-dimensional subspaces. We take advantage of this transformation, which allows us to trade the exponential complexity of the starting problem with a polynomial one. 

The same change of basis simplifies the study even in the presence of dissipation. If we consider the eigenvalue problem related to Eq.~\eqref{eq_def_liouvillian} in momentum space, we get
\begin{equation}
\mathcal{L}[\rho]=\sum_{k>0}\mathcal{L}_k[\rho_k]\,,\quad \mathcal{L}_k[\rho_{k}]=\beta^{(j)}_{k}\rho^{(j)}_{k}\,,
\label{eq_mathcal_L_k}
\end{equation}
where $\rho=\bigotimes_{k>0}\rho_k$ and the superoperator $\mathcal{L}_k$ reads as
\begin{equation}
\begin{aligned}
        \mathcal{L}_k[\rho_k]=&-i[H_k,\rho_k]+w\hat{c}_{k}\rho_k\hat{c}_{k}^{\dagger}-\frac{w}{2}\{c^\dagger_{k}\hat{c}_{k}, \rho_k\}\\
        &+w\hat{c}_{-k}\rho_k\hat{c}_{-k}^{\dagger}-\frac{w}{2}\{c^\dagger_{-k}\hat{c}_{-k}, \rho_k\}\,.
\end{aligned}
\label{eq_app_def_mathcalL_momentum}
\end{equation}
In eq.~\eqref{eq_mathcal_L_k}, the complex number $\beta^{(j)}_{k}\in\mathbb{C}$ denotes the $j$-th eigenvalue associated with the $k$-th Hilbert space, so that $\lambda_r$ are the eigenvalues of $\mathcal{L}$ that are fully specified by $\lambda_r=\sum_k\beta_k^{(a_k)}$ with $a_k=1,\dots,16$. Within each momentum sector, the $16$ eigenvalues $\beta^{(j)}_k$ are explicitly given by
\begin{equation}
    \beta^{(j)}_k=
    \begin{cases}
        0\\
        -w \quad &\text{deg. 4}\\
        -w/2\pm\sqrt{-4-\mu^2-4\mu\cos{k}}\quad &\text{deg. 2}\\
        -w\pm2\sqrt{-4-\mu^2-4\mu\cos{k}}\\
        -3w/2\pm\sqrt{-4-\mu^2-4\mu\cos{k}}\quad &\text{deg. 2}\\   
        -2w\,,
    \end{cases}
\end{equation}
where on the right side we indicate the degeneracy of each eigenvalue. It is now simple to show that the Liouville gap is always equal to 
\begin{equation}
\Delta_\lambda=\frac{w}{2}\,,
\label{eq_liouville_gap_w_2}
\end{equation}
independently of the chemical potential $\mu$ considered. The NESS is the only matrix $\rho^{(0)}$ surviving at asymptotically large times; it satisfies $\mathcal{L}_k[\rho^{(0)}_k]=0$ for all $k>0$. We remark that the existence and uniqueness of a steady-state solution, in general terms, cannot be taken for granted~\cite{N-2019-uniquenesslindblad}. Nonetheless, we were able to find a closed-form expression for this state within each Hilbert domain $\mathcal{H}_{k}$
\begin{equation}
\scriptsize{
\rho^{(0)}_k=
\begin{pmatrix}
    1-\frac{3(1-\cos(2k))}{2g_k(\mu,w)} & 0 &0&
    \frac{|\sin k|(2\mu+i w+4\cos k)}{2g_k(\mu,w)}\\
    0 & \frac{\sin^2k}{g_k(\mu,w)} &0&0\\
    0 & 0&\frac{\sin^2k}{g_k(\mu,w)} &0\\
        \frac{|\sin k|(2\mu-i w+4\cos k)}{2g_k(\mu,w)} & 0&0&\frac{\sin^2k}{g_k(\mu,w)}
\end{pmatrix}
}
\end{equation}
where $g_k(\mu,w)=4+\mu^2 + w^2/4 + 4\mu \cos 
k$. Even if the system is coupled with particle-decay operators that continuously remove particles from the ring, the NESS can exhibit a non-vanishing density of fermions—the total number of particles is not preserved by $\hat{H}$. For instance, the average number of particles per site in the asymptotic limit $t\to+\infty$ is
\begin{equation}
    \frac{1}{L}\sum_x\langle\hat{n}_x\rangle=\frac{4}{L}\sum_{n=1}^{L/2}\frac{\sin^2\big[\frac{(2n-1)\pi}{L}\big]}{4+\mu^2+\frac{w^2}{4}+4\mu\cos[\frac{(2n-1)\pi}{L}]}\,.
\end{equation}
We verified numerically the above equation. 

\section{Simulation techniques}
\label{sec_app_simulation}
In this Appendix, we summarize the numerical techniques employed for the real-time evolution of the Kitaev ring in Eq.~\eqref{eq_def_dissipator} and the determination of the gap $\Delta_\lambda$~\cite{P-2008-thirdquantization}. We also review the Kitaev model investigated in this paper in momentum space for $b\geq1$.

\subsection{Time evolution of two-point functions (coordinate space)}
\label{sec_app_coordinate_space}

The algorithmic details related to the real-time evolution of correlation functions are thoroughly explained in Refs.~\cite{NRV-2019-competingdissipativeandcoherent, TV-2021-dissipativeboundaries}. We generalize these techniques to the dissipation mechanism described by Eq.~\eqref{eq_def_liouvillian}. We adopt these numerical techniques every time the number of local dissipators $n$ is fixed. Essentially, we find a closed system of coupled differential equations allowing us to describe the time evolution of the two-point functions $C(x, y, t)=C_{x,y}$ and $P(x, y, t)=P_{x,y}$ defined in Eq.~\eqref{eq_def_two_point_functions_C_P}. Differentiating these observables with respect to time, we obtain the following differential equations
\begin{eqnarray}
&&  \frac{d}{dt}\,{C}_{x,y} = i\,({C}_{x,y+1} -
  {C}_{x-1,y} + {C}_{x,y-1} - {C}_{x+1,y})
  \qquad \label{eqscxy}\\
&&\quad -i \, ({P}_{y,x-1}^\dagger
  - {P}_{y,x+1}^\dagger - {P}_{x,y-1} + {P}_{x,y+1})
  \nonumber \\
  & &  \quad
  - \frac{w}{2} \sum_{\overset{j=1}{\text{(mod b)}}}^n \,( \delta_{j,y}\,
  {C}_{x,j} + \delta_{j,x}\, {C}_{j,y} 
+\delta_{j,y}\, {C}_{x,j} + \delta _{1,j}\,
  {C}_{j,y}) \nonumber\\
&&\quad  \,,\nonumber\\
&&\frac{d}{dt}\,{P}_{x,y} = -i\,({P}_{x,y+1} +
    {P}_{x+1,y}+ {P}_{x,y-1} + {P}_{x-1,y})
    \nonumber\\
 && \quad -
i\,( {C}_{x,y-1} -
{C}_{y,x-1} - {C}_{x,y+1} 
+ {C}_{y,x+1}) \nonumber \\
 && \quad -
i\,(\delta_{x-1,y} - \delta_{x+1,y})
- 2\,i\,\mu \,{P}_{x,y} \nonumber \\
&&\quad - \frac{w}{2} \sum_{\overset{j=1}{\text{(mod b)}}}^n\,(\delta_{j,y}\,{P}_{x,j} +
\delta_{j,x}\,{P}_{j,y}+ \delta _{j,y}\,{P}_{x,j}
+ \delta_{j,x}\,{P}_{j,y}) \,.  
\nonumber
\end{eqnarray}
We then use standard $4^{\text{th}}$-order Runge-Kutta techniques to solve this system, using the initial conditions given by
\begin{align}
    C_{x,y}(0)&=\Tr[\rho(0)(\hat{c}^\dagger_x\hat{c}_y+\hat{c}^\dagger_y\hat{c}_x)]\\
    P_{x,y}(0)&=\Tr[\rho(0)(\hat{c}^\dagger_x\hat{c}^\dagger_y+\hat{c}_y\hat{c}_x)]\,.
    \label{eq_operators_CP_in_appendix}
\end{align}

\subsection{Time evolution for $b\geq1$ (momentum space)}
\label{sec_app_momentum_space}

Following the quench protocol outlined in Sec.~\ref{sec_observables}, we first set the starting density matrix $\rho(0)$ to the ground state of $\hat{H}$ from Eq.~\eqref{eq_app_Hk_b1}. In particular, the mixture $\rho(0)$ can be rewritten in terms of the reduced density matrices defined within each $k$-sector $\rho_k(0)=\ket{\Omega_k}\bra{\Omega_k}$ as
\begin{equation}
    \rho(0)=\bigotimes_{k>0}\rho_k(0)\,.
\end{equation}
For $t>0$, unfortunately, $\rho_k(t)$ does not remain inside the same $k$-sector as $\rho_k(0)$ (unless $b=1$), and we are forced to study the time evolution of $\rho_k(t)$ in larger domains. For this reason, we define $n/2$ orthogonal Hilbert subspaces $\mathcal{H}_{k_a}$, whose dimension is $4^b$, generated by
\begin{equation}
\mathcal{H}_{k_a}=\text{span}\bigotimes_{m=1}^b\{\ket{0_{k_{a}^m}}, \ket{1_{k_{a}^m}}, \ket{1_{-k_{a}^m}}, \ket{1_{k_{a}^m,-k_{a}^m}}\}\,,
\end{equation}
where $k^m_a\equiv k_a+2\pi m/b$ and $k_a=\pi(2a-1)/L$ with $a=1,\dots,n/2$. We then express $\rho(t)$ as $\rho=\bigotimes_{a=1}^{n/2}\rho_{k_a}$, where $\rho_{k_a}$ is a reduced density matrix living entirely in $\mathcal{H}_{k_a}$. The time evolution of each $\rho_{k_a}$ is controlled by a self-contained Lindblad equation
\begin{equation}
    \frac{d\rho_{k_a}}{dt}=-i[\hat{H}_{k_a},\rho_{k_a}]+w\mathbb{D}[\rho_{k_a}]\,,
\end{equation}
where $\hat{H}_{k_a}=\sum_{m=1}^b\hat{H}_{k^m_a}$. Straightforward manipulations allow us to write the dissipator $\mathbb{D}[\rho_{k_a}]$ in the following form
\begin{equation}
\begin{aligned}
    \mathbb{D}[\rho_{k_a}]&=\frac{1}{b}\sum_{m,p=1}^b\bigg(\hat{c}_{k_a^m}\rho_{k_a}\hat{c}^\dagger_{k_a^{p}}-\frac{1}{2}\{\hat{c}^\dagger_{k_a^m}\hat{c}_{k_a^{p}},\rho_{k_a}\}\\
    &+\hat{c}_{-{k_a^m}}\rho_{k_a}\hat{c}^\dagger_{-k_a^{p}}-\frac{1}{2}\{\hat{c}^\dagger_{-{k_a^m}}\hat{c}_{-k_a^{p}},\rho_{k_a}\}\bigg)\,;
    \label{eq_momentum_space_dissipation_1_over_b}
\end{aligned}
\end{equation}
note that a prefactor $1/b$ naturally emerges in this context. 
Now, the two-point function expectation values, such as $C(x,y,t)$ or $P(x, y, t)$ in Eq.~\eqref{eq_def_two_point_functions_C_P}, can be evaluated directly. For instance, $C(x,y,t)$ takes the form
\begin{equation}
    C(x,y,t)=\frac{2}{L}\Re{\sum_{m, a, q}e^{\mp ik_a^m x}e^{iqy}\langle\hat{c}^\dagger_{\pm k_a^m}\hat{c}_q\rangle}\,.
\end{equation}
The above equation can be further simplified since $\langle\hat{c}^\dagger_{\pm k_a^m}\hat{c}_q\rangle$, where $\langle\hat{c}^\dagger_{k}\hat{c}_q\rangle\equiv \text{Tr}[\rho(t)\hat{c}^\dagger_k\hat{c}_q]$, is non vanishing if and only if $q_a=\pm k^{p}_a$ for $p=1,\dots,b$. Thus we obtain
\begin{equation}
\begin{aligned}
    C(x,y,t)=\frac{2}{L}&\sum_{a=1}^{n/2}\Re\bigg\{\sum_{m,p=1}^be^{\mp ik_a^mx}e^{ik^{p}_a y}\langle\hat{c}^\dagger_{\pm k_a^m}\hat{c}_{k^{p}_a}\rangle\\
    &+\sum_{m,p=1}^be^{\mp ik_a^m x}e^{-ik^{p}_a y}\langle\hat{c}^\dagger_{\pm k_a^m}\hat{c}_{-k^{p}_a}\rangle\bigg\}\,.
\end{aligned}
\end{equation}
Analogous equations can be obtained for different two-point functions as well with similar manipulations. We finally provide an explicit expression for the entanglement entropy $S(t)$ that can be easily expressed within each sector $\mathcal{H}_{k_a}$ as
\begin{equation}
S=-\sum_{a=1}^{n/2}\sum_{m=1}^{4^b}\lambda_{a,m}\log\lambda_{a,m}\,,
\label{eq_def_entropies_momentum_space}
\end{equation}
where $\lambda_{a,m}$ is the $m$-th eigenvalue of the reduced density matrix $\rho_{k_a}$. 

\subsection{Third quantization techniques: coordinate basis}
\label{sec_app_thirdquant_coordspace}

We use third-quantization techniques presented in Ref.~\cite{P-2008-thirdquantization} to compute the Liouvillian gap of the Kitaev rings considered in this work. The results shown in this paper have been obtained from the diagonalization of $4L \times 4L$ antisymmetric complex matrix $A$ defined as
\begin{equation}
\begin{aligned}
    A_{2j-1, 2k-1}=&-2iH_{jk}-D_{jk}/2+D_{kj}/2\\
    A_{2j-1, 2k}=&iD_{kj}\\
    A_{2j, 2k-1}=&-iD_{jk}\\
    A_{2j, 2k}=&-2iH_{jk}+D_{jk}/2-D_{kj}/2\,.
    \label{eq_def_A_matrix_appendix}
\end{aligned}
\end{equation}
The $2L \times 2L$ matrices $H_{jk}$ and $D_{jk}$ are determined, respectively, by the hamiltonian and dissipation processes written in terms of Majorana fermion operators $\{\hat{v}_j\}$, which are defined as
\begin{equation}
    \hat{v}_{2j-2}=(\hat{c}_j + \hat{c}^\dagger_j)\,,\quad \hat{v}_{2j-1}=i(\hat{c}_j-\hat{c}^\dagger_j)\,.
\end{equation}
The Hamiltonian matrix $H_{jk}$ reads as follows (here the indices range from $0$ up to $2L-1$)
\begin{equation}
\begin{aligned}
\hat{H}=&\sum_{jk}\hat{v}_jH_{jk}\hat{v}_k=\frac{1}{4}\sum_{j=0}^{L-1}\big(-2i\hat{v}_{2j+1}\hat{v}_{2j+2}
\\
&+i\mu\hat{v}_{2j}\hat{v}_{2j+1}+h.c.\big)\,,
\end{aligned}
\label{eq_hamiltonian_third_quantization_coordinate}
\end{equation}
where due to APBC we have $\hat{v}_{x+2L}=-\hat{v}_{2x}$. The matrix elements of $D_{jk}$ are instead given by
\begin{equation}
\begin{aligned}
   D_{jk} &= \frac{w}{4}\sum_{k=0}^{n-1}\big(\delta_{2bk,2bk}+i\delta_{2bk, 2bk+1}\\-&i\delta_{2bk+1,2bk}+\delta_{2bk+1,2bk+1}\big)
\end{aligned}
\label{eq_dissipation_third_quantization_coordinate}
\end{equation}
If $\beta_j$ are the eigenvalues of the matrix $A$, known as the \textit{rapitidies}~\cite{P-2008-thirdquantization}, all eigenvalues come in pairs $\beta_j, -\beta_j$ due to the algebraic properties of complex antisymmetric matrices. The Liouvillian gap is finally given by
\begin{equation}
    \Delta_\lambda=2\min_j[\abs{\Re \beta_j}]
\end{equation}

\subsection{Third quantization techniques: momentum basis}
\label{sec_app_third_quantization_momentum_basis}

We use third-quantization techniques also in momentum space to facilitate the evaluation of the Liouvillian gap $\Delta_\lambda$. This strategy is surely convenient for moderate values of $b$, since one trades the diagonalization of a unique $4L \times 4L$ matrix with the diagonalization of $n/2$ matrices $A^a$ of dimension $8b \times 8b$.

We thus define the antisymmetric matrix $A^a$ corresponding to the momentum $k_a=\pi(2a-1)/L$ with $a=1,\dots,n/2$ as
\begin{equation}
\begin{aligned}
    A^a_{2j,2k}=&-2i H^a_{jk}+D^a_{kj}/2-D^a_{jk}/2\\
    A^a_{2j,2k+1}=&iD^a_{jk}\\
    A^a_{2j+1,2k}=&-iD^a_{kj}\\
    A^a_{2j+1,2k+1}=&-2iH^a_{jk}+D^a_{jk}/2-D^a_{kj}/2&\,,
\end{aligned}
\label{eq_A_third_quantization_momentum_basis}
\end{equation}
where $H^a_{jk}$ and $D^a_{jk}$ are $4^b \times 4^b$ matrices that correspond, respectively, to the hamiltonian and dissipation operators acting on the Hilbert space $\mathcal{H}_{k_a}$. After introducing Majorana fermions, one obtains for $H^a_{jk}$
\begin{equation}
\begin{aligned}
    H^a=&\frac{1}{4}\sum_{p=0}^{2b-1}\big(i\delta_{2p,2p+1}+2i\cos{k_a^p}\delta_{2p,2p+1}+h.c.\big)\\
    +&\frac{1}{2}\sum_{p=0}^{b-1}\abs{\sin{k_a^p}}\big(i\delta_{2p,2b+2p+1}+i\delta_{2p+1,2b+2p}+h.c.\big)\,,
\end{aligned}
\label{eq_def_Hamiltonian_third_quantization_momentum_space}
\end{equation}
where we recall that $k^p_a=\pi(2a-1)/L+2\pi p/b$. The matrix elements of the dissipation matrix $D^a_{jk}$ are instead given by
\begin{equation}
\begin{aligned}
    D^a=&\frac{1}{4}\sum_{p=0}^{b-1}\sum_{q=0}^{b-1}\big(\delta_{2\omega_a(p),2\omega_a(q)}+\delta_{2\omega_a(p),2\omega_a(q)+1}\\
-&\delta_{2\omega_a(p)+1,2\omega_a(q)}+\delta_{2\omega_a(p)+1,2\omega_a(q)+1}\big)\,,
\end{aligned}
    \label{eq_dissipation_third_quantization_momentum}
\end{equation}
where $\omega_a(p)$ is a shorthand notation standing for
\begin{equation}
    \omega_a(p)=
    \begin{cases}
    b + p \quad &\text{if} \ k_a^p\geq\pi\\
    p \quad &\text{if} \ k_a^p<\pi\,.
    \end{cases}
\label{eq_third_quantization_momentum_omega_a_p}
\end{equation}
Again, if $\beta^a_j$ are the eigenvalues of the matrices $A^a$, the Liouvillian gap is then given by
\begin{equation}
    \Delta_\lambda=2\min_{a,j}[\abs{\Re \beta^a_j}]\,,
    \label{eq_third_quantization_def_beta_j_a}
\end{equation}
since all rapitidies $\beta_j$ always come in pairs $\beta_j,-\beta_j$.

\nocite{CPR-2022-otto_engine, NCSD-2023-sshmodeltop, FHK-2022-universalfluc, ZWC-2022-exponentialanderson, AC-2020-spread_corr, CKV-2022-zenocrossover}

\bibliography{references.bib}

\end{document}